\documentstyle[12pt,axodraw,epsfig]{article}

\begin{document}

\begin{center} 
{\Large {\bf History of Supersymmetric Extensions 
\bigskip \\
of the Standard Model}}
\end{center}

\vspace{5mm}

\begin{center}
M. C. Rodriguez$^a$ \\
$^a${\it Universidade Federal do Rio Grande - FURG \\
Instituto de Matem\'atica, Estat\'\i stica e F\'\i sica - IMEF \\
Av. It\'alia, km 8, Campus Carreiros \\
96201-900, Rio Grande, RS \\
Brazil}
\vspace{5mm}
\\
\end{center}

\date{\today}


\begin{abstract}
\vspace{3mm}
We recall the many obstacles which seemed, long ago, to prevent
supersymmetry from possibly being a fundamental symmetry of Nature. We also 
present their solutions, leading to the construction of the supersymmetric extensions 
of the Standard Model. Finally we discuss briefly the early experimental searches for 
Supersymmetry.
\end{abstract}


\vspace{5mm}

\section{Introduction}

Although the Standard Model (SM) \cite{sg} based on the gauge symmetry \linebreak
$SU(3)_{c}\otimes SU(2)_{L}\otimes U(1)_{Y}$ describes the observed properties
of charged leptons and quarks it is not the ultimate theory. 
The necessity to go beyond it, from the 
experimental point of view, comes at the moment only from neutrino 
data.  If neutrinos are massive, and oscillate, new physics beyond the SM is needed.

Supersymmetry (SUSY) is an extremely interesting mathematical structure which 
arose in theoretical papers more than 30 years ago independently by Golfand and Likhtman~\cite{gl},
Volkov and Akulov~\cite{va} and Wess and Zumino~\cite{wz}.
The  supersymmetry algebra was introduced in~\cite{gl} \footnote{They also constructed the first 
four-dimensional field theory with supersymmetry, (massive) quantum electrodynamics of 
spinors and scalars}. There 
and in the Wess-Zumino article~\cite{wz}, the supersymmetry generator $Q$ relates 
bosons with fermions in the usual sense. The Volkov-Akulov article, however, deals only 
with fermions. The supersymmetry generator 
acts in a non-linear way, turning a fermion field into a composite bosonic one made of 
two fermion fields~\cite{va} \footnote{They started the foundations of supergravity}. This illustrates 
that the supersymmetric algebra by itself does not 
require superpartners -- in contrast with what is commonly said or thought now.

Since that time many papers have appeared.  This remarkable activity is due to the unique mathematical
nature of supersymmetric theories, which provide possible solutions for various
problems of the SM within its supersymmetric extensions, 
opening the perspective for a
unification of all interactions in the framework of a single
theory~\cite{ssm,susy,wb,dress,tata}.

There are no direct indications on existence of supersymmetry in particle
physics, however there are a number of theoretical and phenomenological issues that the SM fails to address 
adequately:
\begin{itemize}
\item Unification with gravity; The point is that SUSY algebra being a generalization of
Poincar\'e algebra \cite{wb,dress,tata,grav}:
\begin{eqnarray} 
\left\{ Q_\alpha , Q_\beta \right\} &=& \left\{ \overline{Q}_{\dot{\alpha}} ,
\overline{Q}_{\dot{\beta}} \right\} = 0; \nonumber \\
\left\{ Q_\alpha ,  \overline{Q}_{\dot{\beta}} \right\} &=& 2
\sigma_{\alpha \dot{\beta}}^{m} P_{m}; \ \ \ \ \ \ \ \
\left[Q_{\alpha}, P_{\mu} \right] = 0. 
\label{e12b}
\end{eqnarray}
Therefore, an anticommutator of two SUSY transformations is a local coordinate translation. And
a theory which is invariant under the general coordinate transformation is General
Relativity. Thus, making SUSY local, one obtains General Relativity, or a theory of
gravity, or supergravity~\cite{sugra}.
\item Unification of Gauge Couplings; According to {\em hypothesis} of Grand Unification Theory (GUT) all gauge couplings change with energy. All known interactions are the branches of a single
interaction associated with a simple gauge group which includes the group of the SM. To reach 
this goal one has to examine how the coupling change with energy. Considerating the evolution 
of the inverse couplings, one can see that in the SM unification of the gauge
couplings is impossible. In the supersymmetric case the slopes of Renormalization Group Equation curves are changed and the results 
show that in supersymmetric model one can achieve perfect unification \cite{ABF}.
\item Hierarchy problem; The supersymmetry automatically cancels all
quadratic corrections in all orders of perturbation theory due to
the contributions of superpartners of the ordinary particles. The
contributions of the boson loops are cancelled by those of fermions
due to additional factor $(-1)$ coming from Fermi statistic. This
cancellation is true up to the SUSY breaking scale, $M_{SUSY}$,
since
\begin{equation}
 \sum_{bosons} m^2 - \sum_{fermions} m^2= M_{SUSY}^2,
\end{equation}
which  should not be very large ($\leq$ 1 TeV) to make the fine-tuning natural. 
Therefore, it provides a solution to the hierarchy 
problem by protecting the electroweak scale from large radiative corrections 
\cite{INO82a}. However, the origin of the hierarchy is  the other part of the problem. We show below how SUSY can
explain this part as well.
\item Electroweak symmetry breaking (EWSB); The ``running" of the Higgs
masses leads to the phenomenon known as 
{\em radiative electroweak symmetry breaking}. Indeed, the mass parameters from the Higgs
potential $m_1^2$ and $m_2^2$ (or one of them) decrease while
running from the GUT scale to the scale $M_Z$ may even change the
sign. As a result for some value of the momentum $Q^2$ the potential
may acquire a nontrivial minimum. This triggers spontaneous breaking
of $SU(2)$ symmetry. The vacuum expectations of the Higgs fields
acquire nonzero values and provide masses to fermions and gauge bosons, 
and additional masses to their superpartners \cite{running}. Thus the breaking of the electroweak symmetry is not introduced by brute force as in the SM, but appears naturally from the radiative corrections.
\end{itemize}

However SUSY seemed, in the early days, clearly inappropriate 
for a description of our physical world, for obvious and less obvious reasons, 
which often tend to be forgotten, 
now that we got so accustomed to deal with supersymmetric 
extensions of the SM.  Following \cite{Fayet:2001xk}, on this article, we first we recall the obstacles which seemed, long ago, to prevent
SUSY from possibly being a fundamental symmetry of Nature. 

Starting in the early 1980's, people began to realize that SUSY might indeed solve some basic problems 
of our world. Appeared a lot of SUSY correct predictions, some of them are \cite{Chung:2003fi}:
\begin{itemize}
\item  Supersymmetry predicted in the early 1980s that the 
top quark would be heavy \cite{Ibanez:wd}, because this was a   
necessary condition for the validity of the electroweak 
symmetry breaking explanation.

\item Supersymmetric grand unified theories with a high fundamental
scale accurately predicted the present experimental value of 
$\sin^{2}\theta _{W}$ before it was measured
\cite{Dimopoulos:1981yj}.

\item Supersymmetry requires a light Higgs boson to exist
\cite{Kane:1992kq}, consistent with current precision
measurements, which suggest $M_{h} < 200$ GeV \cite{lepewwg}.
\end{itemize}
Together these successes provide powerful indirect evidence that low energy 
SUSY is indeed part of correct description of nature. Our second goal is to present 
a review about the construction of the Supersymmetrics extensions of the SM, as well we present 
the earliest experimental search to SUSY.

The paper is organized as follows. In Sec.~\ref{sec:prob} we present the main difficulties in 
constructing phenomenological supersymmetric extensions of the SM, with their solutions in Secs.~\ref{sec:hrcmssm} 
and \ref{sec:ovn}. Sec.~\ref{sec:ess} is a short review of early experimental searches for SUSY. 
Appendices deal with $R$-symmetry and Feynman rules in the strong sector of 
supersymmetric extensions of the Standard Model.

\section{Nature does not seem to be supersymmetric!}
\label{sec:prob}

We review here some obstacles which seemed, long ago, to prevent SUSY from being a 
fundamental symmetry of Nature. The first crucial question we can formulate, if SUSY is to 
be relevant in particle physics, is:
\begin{itemize}
\item Which bosons and fermions could be related by SUSY?

There seems to be no answer since wew know more bosons
than fermions. By another hand, they do not appear to have much in common. Maybe, SUSY could act at the level of composite objects, e.g. 
as relating baryons with mesons\,?

We remind that in november 1974 
a new vector meson was discovered independently by two groups in Brookhaven  and Stanford (USA), 
denoted by $J$ or $\Psi$, now generally called $J/ \Psi$. 
The true nature of this particle was the subject of lively debates, with
the final explanation provided by the quark model. The $J/ \Psi$ is a 
bound state of a new quark, $c$ (for charm), with $J/ \Psi \equiv c \bar{c}$. 

In 
the quark model there are more baryons (composed of three quarks with half-integer spin, obeying Fermi-Dirac statistics) 
than mesons (composed of a quark and an antiquark, with integer spin, and obeying Bose-Einstein statistics). 
In the original Eightfold Way proposed by 
Murray Gell-Mann in 1961, there are 18 baryons forming an octet and a decuplet, and 8 mesons. With four quarks $u,d,s$ and $c$, the difference between these numbers gets bigger.

We might try to relate baryons and mesons, but what to do with the baryons ``in excess''? It was thus realized that it 
would be very difficult to relate them through the supersymmetric algebra: how could we deal with the fact that 
there are more baryons than mesons\,?

This leads to ask: Should supersymmetry act at a 
fundamental level, i.e. at the level of leptons, quarks and gauge bosons\,? We can 
try do it, possibly with the help of the idea, tentatively pursued in \cite{R}, that the photon could be related with the neutrino, 
as both are massless (or almost massless) and have no electric charge. 

But how do we know that the solution is like that\,? The answer is in fact that it is not ...
However, if we try to follow this way for a while we arrive to another problem. The known leptons and quarks 
are Dirac fermions carrying conserved baryon number ($B$) and lepton number ($L$). All observed processes in nature respect conservation laws for $B$ and $ L$. 

In the beginning of 1974, only two fermion families were known and not even complete 
with the charm quark still to be discovered.
Neutral current effects had just been discovered the year before, in 1973,
with very little information available
about the structure of the weak neutral current (or currents\,?). And the 
lower limit on the mass of the postulated charged $\,W\,$ boson
was something like \,5 \,GeV. 
The Standard Model was a recent theoretical construction, far from ``Standard'' yet
in today's sense; its $\,W^\pm$ and $Z$ bosons, of course hypothetical, were considered as really very heavy. And even-more-hypothetical Higgs fields were generally viewed 
as a technical device to trigger or mimic the 
spontaneous breaking of the gauge symmetry.

In addition, one had to face the question:
\item How could one define (conserved) baryon and lepton numbers in a supersymmetric theory ?
This appeared especially difficult as $B$ and $L$  are known to be carried by fundamental
 fermions only, not bosons ...

The translation generator (momentum) $P_{m}$  appears in the supersymmetric algebra, see Eq.(\ref{e12b}), in agreement 
with Lorentz covariance. Therefore, SUSY is a space-time symmetry and it is independent of all internal 
symmetries. This meant thet the generators of supersymmetric transformations commutes with the generator 
of any internal symmetries.

This will lead us to attribute baryon and lepton numbers also to fundamental bosons, now called 
squarks and sleptons, as well as to fundmanetal fermions. Nowadays we are so used to deal with spin-0 squarks and sleptons, 
carrying baryon and lepton numbers almost by definition, 
that we can hardly imagine this could once have appeared as a problem.

Attributing baryon and lepton numbers to bosons as well as to fermions, however, could lead to an immediate disaster, preventing us from 
getting a theory that conserve $B$ and $L$ quantum numbers. This may be formulated through the question:
\item How can we avoid unwanted interactions mediated by spin-0 squark and slepton exchanges?

This problem, in fact, can be avoided thanks to $R$-parity. The $R$-symmetry was introduced in 1975
by P. Fayet \cite{R} within a first $SU(2)\otimes U(1)$ electroweak theory (see also A. Salam and J. Strathdee
in \cite{r1}), then extended to the gauge interactions of quark and lepton superfields, and to their superpotential
interactions with the two doublet Higgs superfields now known as $H_1$ and $H_2$ in \cite{ssm}.
This continuous $R$ symmetry, later reduced to its discrete $R$-parity subgroup, 
allows for $B$ and $L$ conservation laws by forbidding 
unwanted superpotential interactions that would violate either baryon or lepton
number conservation laws (or both). Nice reviews may be found in
\cite{barbier,moreau}. We show in Appendix~\ref{apen:rsymmetry} that the terms
\begin{eqnarray}
   & &   \int d^{4}\theta\;  \ \bar{ \Phi}(x,\theta ,
   \bar{\theta})\ \Phi(x,\theta ,\bar{\theta}), \nonumber \\
   & &   \int d^{4}\theta\;\ \bar{ \Phi}(x,\theta ,
   \bar{\theta})\ e^{V(x,\theta ,\bar{\theta})} \
   \Phi(x,\theta ,\bar{\theta}), \nonumber \\
   & &   \int d^{2}\theta\;\  \prod_{a}\Phi_{a}(x,\theta ,\bar{\theta}),
              \hspace{1cm} \mbox{if  } \sum_{a} n_{a} = 2,
              \label{invrparity}
\end{eqnarray} 
used in the construction of supersymmetric Lagrangian densities, 
are invariant under the continuous $R$ symmetry provided $\,\sum_{a} n_{a} = 2\,$, 
$n_a$ being the $R$-index of the superfield $\,\Phi_a$.

The great success of the Standard Model with its broken $SU(2)_{L} \otimes U(1)_{Y}$ symmetry
has convinced of the usefulness of broken symmetries.
Unfortunately it is not easy to break SUSY spontaneously. 
One problem follows directly from the SUSY algebra itself, which implies 
that the energy is always non-negative definite (at least in global SUSY models). Indeed, 
$$ E = <0| \ H \ |0> $$ and due to the SUSY algebra, see Eq.(\ref{e12b}), we can write
 $$\{Q_{\alpha}, \bar Q_{\dot \beta} \} =
2(\sigma^{m}) _{\alpha \dot \beta}P_{m} .$$ 
As $\hbox{Tr} (\sigma^{m}P_{m}) = 2P_0 ,$
 one gets formally 
$$E = \frac{1}{4} <0| \left( \overline{Q}_{\dot 1} Q_1 + Q_1 \overline{Q}_{\dot 1}+ 
\overline{Q}_{\dot 2}Q_2 + Q_2 \overline{Q}_{\dot 2}
\right) |0> \,\  \geq 0 ,$$
hence $$ E = <0| \ H \ |0> \neq 0 \ \ \ \ \mbox{if \ and \ only \ if} \ \
\ Q_{\alpha}|0> \neq 0 .$$

According to this argument, SUSY would be spontaneously broken, with the vacuum state
not invariant under SUSY (i.e. $Q_{\alpha} |0> \neq 0 $), if and only if the
minimum of the potential is strictly positive, corresponding to a positive energy density. 
But on the other hand {\it a supersymmetric vacuum state must have vanishing energy, 
and is therefore necessarily stable !}

It thus seemed that SUSY could not be spontaneously broken at all, 
which would imply that bosons and fermions be systematically degenerated 
in mass \footnote{Since $P^{2}$ commutes with $Q$ and $\overline{Q}$, see Eq.(\ref{e12b}), the mass of a particle is the same within 
a supermultiplet representation.}. This is clearly not realistic, as there is no boson with the same mass 511 keV as the electron, 
nor with the same mass 106 MeV as the muon, etc,  unless of course SUSY breaking terms are explicitly 
introduced "by hand". This leads to the question:
\item Is spontaneous SUSY breaking possible at all\,?

The answer is yes but it is not so easy. 
Spontaneous breaking of SUSY may be achieved in a way somewhat similar to the
electroweak symmetry  breaking, with a field whose vacuum expectation value (vev) is non-zero and breaks spontaneously the symmetry.

Due to the special character of SUSY, this field should be one (or a combination) of the 
auxiliary $F$ or $D$ field-components of superfields. One must then be able to arrange, and this is the difficult point, 
so that one at least of these auxiliary fields acquires a non-vanishing vev, although they all tend to reach
as much as possible a point at which they would vanish, so as to minimize the energy.

Among possible spontaneous SUSY breaking
mechanisms one distinguishes the $F$ and $D$ ones.

i) Fayet-Iliopoulos ($D$-term) mechanism \cite{Fayet}. \\
 In this case the, the linear $D$-term is added to the Lagrangian
 \begin{equation} 
\Delta {\cal L} = \xi V\vert_{ \theta \theta \bar \theta \bar \theta} \equiv \xi\int
d^4\theta\ V.
\end{equation}
It is gauge and SUSY invariant by itself.

When this mechanism is applied to supersymmetric extensions of the standard model,
the resulting mass spectrum is problematic, due in particular to the following sum rule
 \begin{equation}
\sum_{boson \ states} m^2_i =
\sum_{fermion \ states} m^2_i , \label{sumrule}
 \end{equation}
valid in a $SU(3)\otimes SU(2)\otimes U(1)$ gauge theory (at the classical level), and in particular in the quark subsector \cite{fayet79}. This would prevent one to make all squarks (and sleptons) heavy, at the classical level,
unless one extends the gauge group to include an extra $U(1)$ factor beyond $SU(3)\otimes SU(2)\otimes U(1)$.

ii) $F$-breaking mechanism, due to Fayet \cite{fayetF} and O'Raifeartaigh \cite{O'R}. \\
Several chiral superfields are needed for that purpose and the superpotential should be
chosen very carefully, with additional restrictions as generally provided by an $R$ symmetry, to avoid the 
otherwise quasi-systematic presence of supersymmetric vacuum states in which all auxiliary components would vanish.
 For instance, choosing the superpotential \cite{O'R}
 $${\cal W}(\Phi)=
\lambda \Phi_3 +m\Phi_1\Phi_2 +g \Phi_3\Phi_1^2,$$
 one gets the equations for the auxiliary fields
 \begin{eqnarray*}
F^*_1&=&mA_2+2gA_1A_3, \\
F^*_2 &=& mA_1, \\
F^*_3 &=& \lambda +gA^2_1,
 \end{eqnarray*}
which have no solution with $<F_i> =0$, so that SUSY is spontaneously broken.

This mechanism necessitates the introduction of a completely new sector for
the breaking of the supersymmetry, which leads a lot of arbitrariness in the
superpotential and resulting potential. The sum rule (\ref{sumrule}) is also valid
here (up to radiative corrections).

Unfortunately, none of these mechanisms may be used directly in a satisfactory way in the minimal versions
of SUSY extensions of the SM, owing also to the difficulty of generating spontaneously sufficiently large gluino masses in these models  \cite{fayet79,glu}. 
Until today, spontaneous SUSY breaking leading to an acceptable mass spectrum remains, 
in general, rather difficult to obtain at least within global supersymmetry.  
We thus generally parametrize our ignorance about the 
true mechanism of SUSY breaking chosen by Nature to make superpartners heavy by introducing in the Lagrangian density additional terms breaking the supersymmetry explicitly, but  {\em softly}, 
in the sense that they do not reintroduce quadratic divergencies in the theory. These terms are 
\cite{10}
\begin{itemize}
\item gaugino mass terms $- \frac{1}{2} M_{a} \lambda_{a} \lambda_{a}$, where $a$ is the group index;
\item scalar mass terms $-M_{\phi_i}^2|\phi_i|^2$; 
\item trilinear scalar interactions $A_{ijk}\, \phi_{i}\phi_{j}\phi_{k}$; 
\item and bilinear terms $-B_{ij}\, \phi_{i}\phi_{j} + h.c.$\footnote{Linear terms $-C_{i} \phi_{i}$ are also allowed, 
where $\phi_i$ is a gauge singlet field.}. 
\end{itemize}

Of course just accepting the possibility of explicit SUSY breaking 
without worrying too much about the origin of 
SUSY breaking terms makes things much easier
\,-- but also at the price of introducing a large number 
of arbitrary parameters, 
coefficients of these supersymmetry-breaking terms. If we allow 
all the new parameters introduced above to be complex in the 
Minimal Supersymmetric Standard Model we would be dealing with 
some 124 unknown real constants, 
19 of which are already in the SM while 105 are new.

Before introducing these soft terms and deal in this way with 
supersymmetry breaking, there was another related question:
\item Where is the spin-$\frac{1}{2}$  Goldstone fermion of SUSY\,?
\end{itemize}
Could it be one of the known neutrinos~\cite{va}\,? If the Goldstone fermion of SUSY is not one of them, why hasn't it been observed\,? Today we tend not to think at all about this question because the generalized 
use of soft terms breaking explicitly the SUSY seems to make this question irrelevant.

\section{Historical review of the construction  of the Minimal Supersymmetric Standard Model}
\label{sec:hrcmssm}

The first attempt to construct a supersymmetric electroweak model of ``leptons'' was done in
\cite{R}, where the author tried to relate known particles together -- in 
particular, the photon with a ``neutrino", and the $W^{\pm}$'s with charged 
``leptons", also related with charged Higgs bosons (now $H^{\pm}$).
This first $SU(2) \otimes U(1)$ electroweak theory involved two doublet Higgs
superfields now known as $H_{1}$ and $H_{2}$~\footnote{Then called $S \equiv H_{1}$ 
left-handed and $T^{c}\equiv H_{2}$ right-handed 
with vev's fixed by $v^{\prime\prime}=v_1$  and $v^{\prime}=v_2$, respectively. See Table \ref{tab:mssm}.}, 
and remains today as the heart of  supersymmetric extensions of the Standard Model \cite{ssm}.

\vspace{1mm}

It describes the $SU(2) \otimes U(1)$ massive gauge bosons $W^{\pm},\,Z^{0}$ and the photon, 
the ``electron'' and its ``neutrino'', together with other heavy particles including a charged scalar 
Higgs boson 
$w^{\pm}$ (now $\,H^\pm$) and a ``heavy electron'' $E^{\pm}$ (see next section 
for more details about the mass 
spectrum). It also introduced $R$-invariance. This continuous $ R$-invariance was 
interpreted at the time as associated with lepton number conservation (quarks 
were not considered in this article), $e^{-},e_{0}$ and $ \nu_{L}$ having ``lepton number'' $+1$, 
\,and $E^{-}$ and $E_{0}$  $-1$.
These charged and neutral ``leptons'' were soon to be reinterpreted as new particles, becoming the ``charginos'' and ``neutralinos'' of supersymmetric extensions of the Standard Model \cite{ssm}.

In 1975 the $R$-invariance were introduced to prevent scalar-particle to be exchanged in the $\mu$-decay, 
it was associated with lepton number conservation. Later, in 1976, it was generalized to also include 
baryonic number conservation. Then the role of the $R$-invariance were to avoid the scalar particle exchange in $\mu$ and $\beta$-decay as 
analysed in 1977 \cite{ssm}. Since  that time, we associate the conservation of $R$-invariance with the conservation of 
lepton and baryon number.

\begin{figure}[tb]
\begin{center}
\epsfig{file=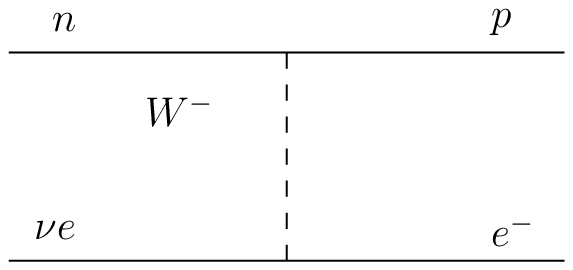,width=4cm}
\epsfig{file=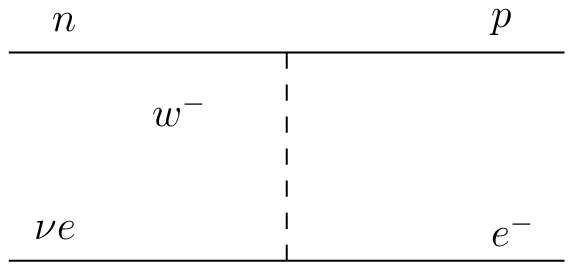,width=4cm}
\end{center}
\caption{Diagrams contributing to $\beta$-decay, taken from the last reference in \cite{ssm}. In the second diagram, a new charged scalar 
Higgs boson $w^{\pm}$ is exchanged. This diagram is negligible with respect to the first since Yukawa coupling constants vanish with fermion masses.}
\label{fig:bdeacy}
\end{figure}

\begin{figure}[tb]
\begin{center}
\epsfig{file=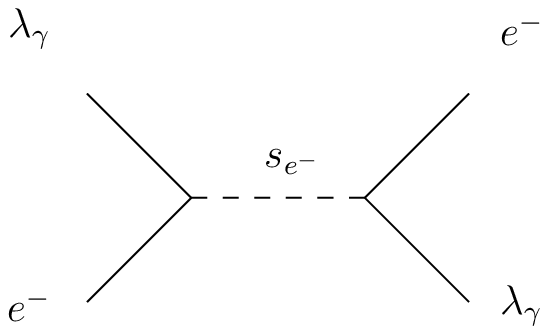,width=4cm}
\epsfig{file=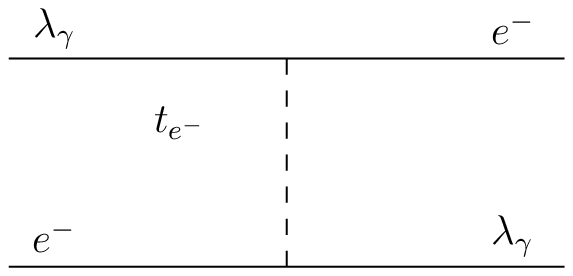,width=4cm}
\end{center}
\caption{Scattering of a new ``neutrino'' $\lambda_{\gamma}$ (now the photino) on electrons induced by the excanges of septons and teptons 
respectivelly, taken from the last reference in \cite{ssm}.}
\label{fig:new}
\end{figure}

Since the new bosons carrying electronic, muonic or baryonic number are all heavy, 
these quantum numbers 
effectively appear as carried by fermions only at low energies. In addition these new spin-0 particles now known as squarks and sleptons cannot be exchanged in $\mu$- or $\beta$- decay (only the $W^\pm$ and charged Higgs being exchanged as in Fig.~\ref{fig:bdeacy}), thanks to $R$-symmetry or simply its discrete version known as $R$-parity. 
Moreover, the scattering 
$\lambda_{\gamma}+e^- \rightarrow \lambda_{\gamma}+e^-$ may be induced by the exchanges of
scalar particles, as shown in Fig.~\ref{fig:new}, where $s_{e^{-}}$ and $t_{e^{-}}$,  first called ``septons" and  ``teptons", became today's sleptons, in this case selectrons; $\lambda_{\gamma}$ being the photino.

Ref.~\cite{ssm} also introduced squarks (called ``sarks" for $s^{f}_{c}$ and 
``tarks" for $t^{f}_{c}$) and gluinos (color octet of Majorana fermions first called ``gluon-neutrinos") denoted by $\lambda_{a}$.  
They couple to squark/quark pairs within what is now known as Supersymmetric Quantum Chromodynamics (SQCD), which also involves quartic interactions for squark fields. 

This article also introduced a second $U(1)$ gauge group with coupling constant $g^{\prime \prime}$, 
originally intended to trigger a spontaneous breaking of the supersymmetry and generate spontaneously
relatively large masses for squarks and sleptons, by inducing $m_\circ^2\,$ squark and slepton mass terms
from the corresponding $<D''>$.
The resulting Goldstone spinor is then a linear combination involving in particular the extra-$U(1)$ gaugino $\lambda^{\prime \prime}$, and the photino $\lambda_{\gamma}$. Mass splittings are obtained, in particular 
for quark and lepton superfields, coupled to $V^{\prime \prime}$
(and to $V_{\gamma}$ for the charged ones).

\begin{figure}[tb]
\begin{center}
\epsfig{file=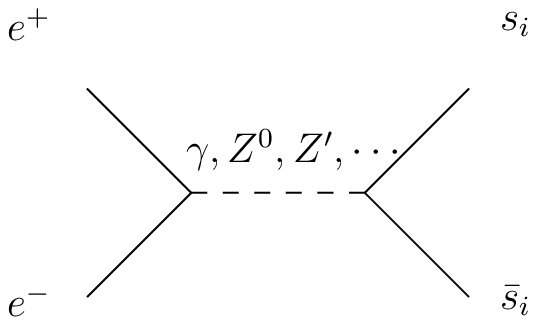,width=4cm}
\epsfig{file=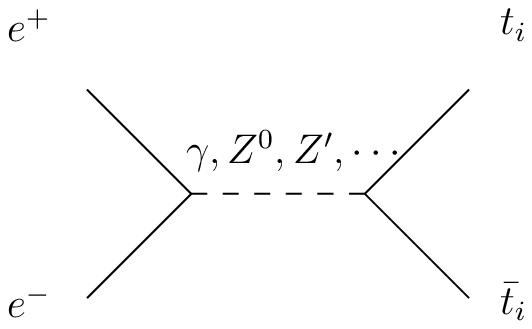,width=4cm}
\epsfig{file=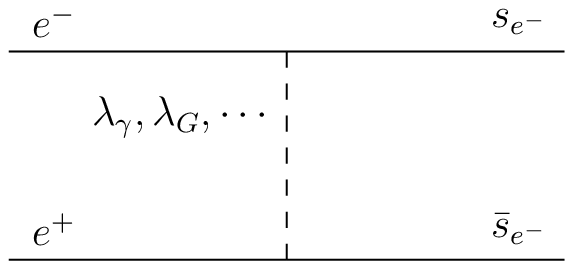,width=4cm}
\epsfig{file=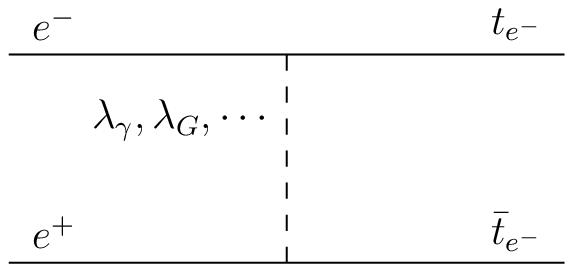,width=4cm}
\end{center}
\caption{Examples of pair-production of new scalars -- now sleptons and squarks -- in $e^{+}e^{-}$ scatterings, taken for the last reference in \cite{ssm}. $\lambda_\gamma$ and $\lambda_G$ denote the photino and goldstino, respectively.}
\label{fig:e-e+}
\end{figure}

\begin{figure}[tb]
\begin{center}
\epsfig{file=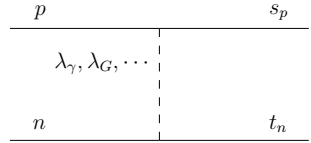,width=4cm}
\end{center}
\caption{Scattering quark + quark into a pair of squarks, taken for the last reference in \cite{ssm}.
$p$ and $n$ stand for the $u$ and $d$ quarks, $s_p$ and $t_n$ for the corresponding squarks $\tilde u$ and $\tilde d$,
$\lambda_\gamma$ and $\lambda_G$ and $\lambda_a$ denote the photino, goldstino and gluinos, respectively.
}
\label{fig:qq}
\end{figure}

The Goldstone neutrino $\lambda_{G}$, the photon-neutrino $\lambda_{\gamma}$, and gluon-neutrino $\lambda_{a}$,
are at this stage massless and carry one unit of $R$-number. The heavy scalars $s_{i}$ and $t_{j}$ associated with quarks and 
leptons carry $R$-number equal $+1$ (for $s_{i}$) or $-1$ (for $t_{j}$). These scalars have very short lifetimes. One of their decay 
modes gives back the corresponding fermion, together with a Goldstone-neutrino, or a photon-neutrino for a charged 
particle. The scalars can be produced in pairs, for example in $e^{+}e^{-}$ scatterings (cf. Fig.~\ref{fig:e-e+}) and 
in quark-quark scatterings as shown in Fig.~\ref{fig:qq}.  

\begin{figure}[tb]
\begin{center}
\epsfig{file=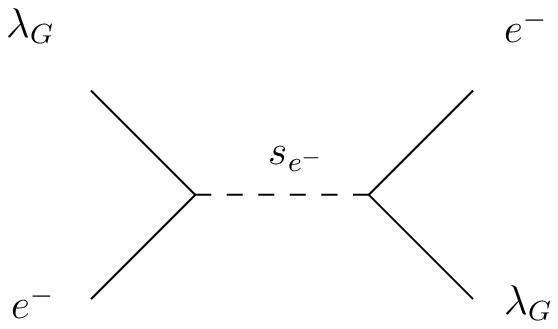,width=4cm}
\epsfig{file=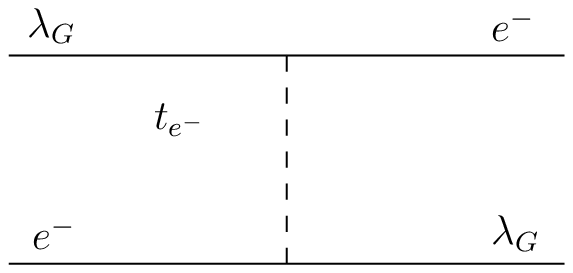,width=4cm}
\epsfig{file=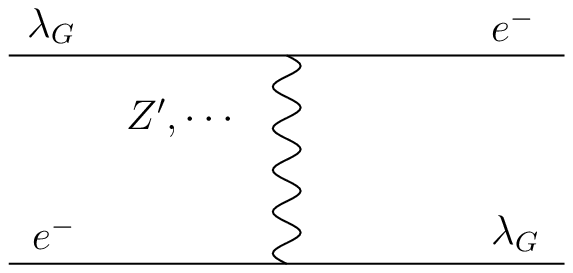,width=4cm}
\end{center}
\caption{Scattering of new neutrino $\lambda_{G}$ on electrons, taken from the last reference in \cite{ssm}.}
\label{fig:lg}
\end{figure}

New processes such as represented in Figs. \ref{fig:new} and \ref{fig:lg} may be important in the study of stellar cooling, 
and supernovae explosions. The process in Fig. \ref{fig:qq} may be relevant for
black holes, or in the very early stages of the Universe. In all reactions described above, the $R$-number is 
conserved. In 1977 the photon-neutrino had its name contracted into photino and gluon-neutrinos into gluinos
\cite{grav}. 
The sarks and tarks, septons and teptons became later the
squarks and sleptons.

However, such an unbroken continuous
$R$-symmetry acting chirally on gauginos, and gluinos in particular, would
maintain the latter massless, even after a spontaneous breaking of the
SUSY \cite{glu}. Indeed the gaugino's
mass term is given by \cite{10} 
\begin{equation} 
m_{\lambda} \left( \lambda \lambda + \bar{\lambda} \bar{\lambda} \right), 
\label{gaugino mass term} 
\end{equation} 
which, under the $R$-symmetry, see Eq.(\ref{The R-Invariance prop 5}), transforms into
\begin{equation} 
m_{\lambda} \left( e^{2i \alpha}\lambda \lambda + e^{-2i \alpha}\bar{\lambda} \bar{\lambda} \right) ,
\end{equation}
so that the mass term (\ref{gaugino mass term}) is not
invariant under $R$. This forces us to abandon the
continuous $R$-symmetry, in favour of its discrete version
called $R$-parity. This one allows for
gluinos and other gauginos to acquire masses.
Moving from $R$-symmetry to $R$-parity is in any case necessary within supergravity, so that the spin-$\frac{3}{2}\,$ (Majorana) gravitino can acquire a mass $\,m_{3/2}$, which does also violate the continuous $R$- symmetry \cite{grav}.

In the early days, though, it was difficult to obtain significant masses for gluinos, 
due the fact that: 
\begin{itemize} 
\item no direct gluino mass term was present in the Lagrangian density; 
\item no such term may be generated spontaneously at the tree 
approximation, since gluino couplings involve colored spin-0 fields which cannot be translated.
\end{itemize}

In this case, gluinos remain massless (at least at the classical level), and we would then expect the existence of relatively light
``$R$-hadrons"\footnote{Particles made of quarks, antiquarks and gluinos, discussed later in Sec.~\ref{sec:ess}.}~\cite{ff,ff2}, 
which have not been observed. We know today that gluinos, 
if they do exist, should be rather heavy, requiring a significant 
breaking of the continuous $\,R$-invariance,
in addition to the necessary breaking of the supersymmetry.

A third reason for abandoning the continuous $\,R$-symmetry in favor of its discrete $R$-parity version 
is now the non-observation at LEP
of a charged wino \,-- also called chargino --\,
lighter than the $\,W^\pm$, \,that would exist 
in the case of a continuous $\,R$-invariance \cite{ssm,R}. 
The just-discovered $\,\tau^-\,$ particle could tentatively be considered, 
in 1976, as a possible light wino/chargino candidate, 
before getting clearly identified as a sequential heavy lepton, the $\tau$ lepton.

Gluino masses may result directly from supergravity, through 
$m_{3/2}$, which also leads to abandon the continuous $R$ symmetry for $R$-parity 
as already observed in 1977~\cite{grav}. Another method, which does not use supergravity, generates $m_{gluino}$ radiatively
using, in modern terminology,  ``messenger quarks'' sensitive to the source of SUSY breaking~\cite{glu}.
The resulting gluino mass, however, would be small, unless the mass of the messenger quarks 
responsable for the generation of gluino masses is taken to be rather large.

The conservation \,(or non-conservation)\,
of $R$-parity is closely related with the conservation 
\,(or non-conservation)\, of baryon and lepton numbers, 
$\,B\,$ and $\,L$, ~as illustrated by the well-known formula 
reexpressing $\,R$-parity in terms of baryon and lepton numbers, 
as $\,(-1)\,^{2S} \ (-1)\,^{3B+L}$
~\cite{ff}. This may also be written as $\ (-1)^{2S} \ (-1)\,^{3\,(B-L)}\,$, 
~showing that this discrete symmetry may still be conserved 
even if baryon and lepton numbers are separately violated,
as long as their difference $\,B-L\,$ remains 
conserved, at least modulo 2.

The finding of the basic building blocks of 
the Supersymmetric Standard Model,
whe\-ther ``minimal'' or not, allowed for the 
experimental searches for ``supersymmetric particles'', 
starting with the first searches for gluinos and photinos, 
selectrons and smuons, in the years 1978-1980,
and going on continuously since.
These searches often rely on the "missing energy'' signature
corresponding to energy-momentum carried away by unobserved 
neutralinos~\cite{ssm,ff,ff2,ff3}.
A conserved $R$-parity also ensures the stability 
of the ``lightest supersymmetric particle'',
\,a good candidate to constitute the non-baryonic Dark Matter 
that seems to be present in the Universe.  

Massive neutrinos can also be naturally 
accommodated in $R$-parity violating supersymmetric theories,
where neutrinos can mix with neutralinos so that they acquire small masses \cite{barbier,Nilles,hall0,GH,dreiner}. 
The
phenomenological bounds
on $B$ and/or $L$ violations \cite{barbier,moreau,dreiner}
can 
be satisfied by
imposing $B$ as a symmetry
and 
allowing $L$-violating couplings
to be sufficient to generate appropriate 
neutrino Majorana masses.

Beyond that, the minimal extension of the MSSM also introduces an extra singlet superfield
now called $S$, this model is called the ``Next to Minimal Supersymmetric Standard Model" (NMSSM) \cite{R,dress}.  

In fact, P. Fayet started in 1975 \cite{R} with a $\mu$ parameter as in the MSSM, corresponding in modern langage to a $\,\mu\,H_1\,H_2$ superpotential term (this $\mu$ was then called $m$). But this $\mu$ parameter immediately turned out to be an obstacle for getting a satisfactory electroweak breaking at tree level, with both Higgs-doublet vev's, now called $\,v_1$ and $v_2$, non-zero. 
He thus modified the theory within the same paper, promoting $\mu$ to the role of
a dynamical superfield variable. I.e. changing it for an extra chiral singlet superfield $S$, or in modern notation, replacing $\mu$ by $\lambda S$. His 
superpotential was initially given by
\begin{equation}
\label{superpot}
\lambda \,H_{1}H_{2}S+f(S)
\end{equation}
as in the ``general NMSSM" with $f(S)$ including in principle $S^{3},S^{2}$ and 
a linear $S$ terms. He then kept only the linear term proportional to $S$ (leading to the model now known as the nMSSM), as he wanted:
\begin{itemize}
\item A continuous $R$-symmetry surviving electroweak (EW) breaking;
\item EW breaking occurring independently of SUSY breaking, i.e. even ``before'' SUSY gets spontaneously broken.
\end{itemize}
Later he also considerated in \cite{ssm} the possibility of gauging an extra $U(1)$ factor in the gauge group (USSM), which requires 
the $f(S)$ terms in the superpotential (\ref{superpot}) to be absent, as $S$ is charged under this extra $U(1)$.

One of the simplest extensions of the SM that 
allows to naturally explain 
the smallness of neutrino masses (without excessively tiny Yukawa
couplings) incorporates right-handed
(Majorana) neutrino fields with a seesaw mechanism \cite{grs79,ms80a} for neutrino
mass generation~\cite{kim,king,Davidson:1998bi,Antusch:2006bw}, it is the ``Minimal Supersymmetric Standard Model with three right-handed neutrinos" (MSSM3RHN) \cite{tata}.

The introduction
of three families of right-handed neutrinos $N^i$ (where $i$ is 
flavor index) brings two new ingredients to the SM;
the Majorana mass scale for the right-handed neutrinos, and a new matrix for their Yukawa coupling constants.
We thus have two independent Yukawa matrices in the lepton
sector as in the quark sector. This model can accommodate a seesaw mechanism,
while stabilising the hierarchy between the scale of new
physics and the EW scale \cite{Teixeira:2007gq}.

\section{The original construction}
\label{sec:ovn}

\subsection{From original notations to modern ones}

Considerating \cite{Fayet:2001xk}, we give an short review on the 
first studies on the MSSM, discussing in particular the connections between 
original notations and the ones used today. The left-handed doublet Higgs superfield $\,S\,$ is now denoted by $\,H_1\,$,
and the right-handed one $T$ is replaced by its conjugate, $\,T^{c}=H_2$.
In the quark and lepton sector the superfields  $\,S_i$, \,left-handed, 
and $\,T_j$, \,right-handed, describe the left-handed and right-handed 
spin-$\frac{1}{2}$ quark and lepton fields, 
together with their spin-0 partners. In today's notation the $\,S_i$'s are the left-handed doublet
superfields $\,L_a$ and $Q_i$, and the $T_j$'s are traded for their conjugates, 
the left-handed singlet superfields $\bar E_a,\ \bar D_i$ and $\,\bar U_i$.

\begin{table}[t]
\center
\renewcommand{\arraystretch}{1.5}
\begin{tabular}
[c]{|l|cc|cc|}\hline
Superfield & Usual Particle & Spin & Superpartner & Spin\\\hline\hline
\quad$V^{\prime}$ (U(1)) & $V_{\mu}$ & 1 & $\lambda^{\prime}\,\,$ & $\frac
{1}{2}$\\
\quad$V^{i}$ (SU(2)) & $V^{i}_{\mu}$ & 1 & $\lambda^{i}$ & $\frac
{1}{2}$\\
\quad$V^{a} (SU(3))$ & $V^{a}_{\mu}$ & 1 & $\lambda_{a}$ &
$\frac{1}{2}$\\\hline
\quad$S_{i}\sim({\bf 3},{\bf 2},1/3)$ & $(u_{i},\,d_{i})_{L}$ & $\frac
{1}{2}$ & $(\tilde{ u}_{iL},\,\tilde{ d}_{iL})$ & 0\\
\quad$T_{u_{i}}\sim({\bf 3^{\ast}},{\bf 1},4/3)$ & ${u}_{iR}$ &
$\frac{1}{2}$ & $\tilde{ u}_{iR}$ & 0\\
\quad$T_{d_{i}}\sim({\bf 3^{\ast}},{\bf 1},-2/3)$ & ${d}_{iR}$ &
$\frac{1}{2}$ & $\tilde{ d}_{iR}$ & 0\\
\quad$S_{i}\sim({\bf 1},{\bf 2},-1)$ & $(\nu_{i},\,l_{i})_{L}$ & $\frac
{1}{2}$ & $(\tilde{ \nu}_{iL},\,\tilde{ l}_{iL})$ & 0\\
\quad$T_{e_{i}}\sim({\bf 1},{\bf 1},-2)$ & ${e}^{-}_{iR}$ &
$\frac{1}{2}$ & $\tilde{ e}^{-}_{iR}$ & 0\\ \hline
\quad$S \sim({\bf 1},{\bf 2},-1) \mbox{(left-h.)}$ & $(S^{0},\, S^{-})$ & 0 &
$(\tilde{ S}^{0},\, \tilde{ S}^{-})$ & $\frac{1}{2}$\\
\quad$T \sim({\bf 1},{\bf 2},1) \mbox{(right-h.)}$ & $(T^{0},\, T^{-})$ & 0 &
$(\tilde{ T}^{0},\, \tilde{ T}^{-})$ & $\frac{1}{2}$\\
\hline
\end{tabular}
\renewcommand{\arraystretch}{1}\caption{Field content of the MSSM following \cite{ssm}.}
\label{tab:mssm}
\end{table}

This model contains the field content indicated in
Table \ref{tab:mssm}. The  family indices are $i,j=1,2,3$. The 
parentheses in the first column refer to transformation properties under $(SU(3)_C,SU(2)_L,U(1)_Y)$.
All this constitutes the basic structure of the 
{\it \,Minimal Supersymmetric Standard Model\,},
which involves the minimal set of ingredients 
shown in Table \ref{tab:basic}.

\begin{table}[t]
\caption{\ The basic ingredients of the MSSM taken from the first reference \cite{Fayet:2001xk}.
\label{tab:basic}}
\vspace{0.2cm}
\begin{center}
\begin{tabular}{|l|} \hline \\ 
1) $\,SU(3)\times SU(2)\times U(1)\,$ gauge superfields; \\ \\
2) \,chiral superfields for  \\
\hskip 1truecm the three quark and lepton families; \\ \\
3) \,two doublet Higgs superfields $\,S$ and $\,T$
 \\ 
\hskip 1truecm responsible for EW breaking,  \\ \\
4) \, quark and lepton masses, through \\
\hskip 1truecm 
the trilinear superpotential given in \,Eq.(\ref{supot})\ .
\\  \\ \hline
\end{tabular}
\end{center}
\end{table}

The Lagrangian of this model was written as
\begin{equation}
{\cal L}^{original}_{MSSM} \equiv {\cal L}_{SUSY}={\cal L}_{lepton}+{\cal L}_{quarks}+{\cal L}_{gauge}+
{\cal L}_{Higgs}+ \xi^{\prime}D^{\prime},
\label{originallagrangian}
\end{equation}
where $\mathcal{L}_{SUSY}$ includes $\xi^{\prime}D^{\prime}$,
the Fayet-Iliopoulos ($D$-term) term for the weak-hypercharge $U(1)$
(its contribution gets today included within the additional ${\cal L}_{soft}$ SUSY-breaking terms). 
The various terms in ${\cal L}_{SUSY}$ are given by 
\begin{eqnarray}
{\cal L}_{lepton}  & =& \int d^{4}\theta\;\left[  \, 
S^{\dagger}_{a}e^{2gV+g^{\prime}F^{\prime} V^{\prime}}S_{a}+
T^{\dagger}_{e_{a}}e^{-g^{\prime}F^{\prime}V^{\prime}}T_{e_{a}}\,\right]
\,\ ,\nonumber\label{The Supersymmetric Term prop 2}\\
{\cal L}_{quarks}  & =& \int d^{4}\theta\;\left[  \,
S^{\dagger}_{i}e^{2g_{s}V_{c}+2gV+g^{\prime}F^{\prime}V^{\prime}}S_{i}+
T^{\dagger}_{u_{i}}e^{-2g_{s}V_{c}-g^{\prime}F^{\prime}V^{\prime}}T_{u_{i}}\right.
\nonumber\\
& +&\left.  
T^{\dagger}_{d_{i}}e^{-2g_{s}V_{c}-g^{\prime}F^{\prime}V^{\prime}}T_{d_{i}}\,\right]  
\,\ ,\nonumber\\
{\cal L}_{gauge}  & =& \frac{1}{4} \left \{ \int d^{2}\theta\;\left[  \sum_{a=1}^{8}W_{s}^{a\alpha}W_{s\alpha}^{a}+
\sum_{i=1}^{3}W^{i\alpha}W_{\alpha}^{i}+W^{\prime\alpha}W_{\alpha}^{\prime} \right]\,+h.c.\right\}
\,, \nonumber \\
\label{mssmlagrangian}
\end{eqnarray}
where the subscripts $a$ and $i$ are family (or group) indices.
These terms correspond to the $\,SU(3)\otimes SU(2)\otimes U(1)$ gauge interactions of quark and lepton superfields.
Before continuing, we remember that SUSY's phenomenology started with the search for
``$R$-hadrons"~\cite{ff,ff2}, which are (unstable) strongly-interacting particles built from gluinos. The 
corresponding Feynman rules 
for the gluinos, quarks and squarks are derived in Appendix~\ref{apen:frsqcd}.

The last piece of the Lagrangian density corresponds to the electroweak interactions of the two doublet Higgs superfields
$S$ and $T$ (now replaced by $\,H_1$ and $H_2$),  written as 
\begin{equation}
{\cal L}_{Higgs} = \int d^{4}\theta\;\left[  \,
S^{\dagger}e^{2gV+g^{\prime}F^{\prime}V^{\prime}}S+
T^{\dagger}e^{-2gV-g^{\prime}F^{\prime}V^{\prime}}T\right]  + 
\int d^{2}\theta \,\ {\cal W}+\int d^{2}\bar{\theta} \,\ \overline{{\cal W}}. \nonumber \\
\label{The Supersymmetric Term prop 4}
\end{equation}
The field strengths are given by \cite{wb}
\begin{eqnarray}
W_{s\alpha}^{a}  & =&-\frac{1}{8g_{s}}\,\bar{D}\bar{D}
e^{-2g_{s}V_{c}^{a}}D_{\alpha}e^{2g_{s}V_{c}^{a}}\,\ \alpha=1,2\,\ ,\nonumber\\
W_{\alpha}^{i}  & =&-\frac{1}{8g}\,\bar{D}\bar{D}e^{-2gV^{i}}D_{\alpha}
e^{2gV^{i}}\,\ ,\nonumber\\
W_{\alpha}^{\prime}  & =&-\frac{1}{4}\,DD\bar{D}_{\alpha}V^{\prime}\,\ .\label{fieldstrength}%
\end{eqnarray}

\subsection{The superpotential for quarks and leptons}

To generate the appropriate quark and lepton mass terms three types of superpotential terms were considered in
\cite{ssm}, generically written as 
\begin{equation}
\label{superpotterms}
S\,T_j^\dagger S_i,\ \ T^\dagger \,T_j^\dagger S_i\ \ \hbox{and} \ \ T_j^\dagger S_i,
\end{equation}
under the condition that they are gauge invariant.
This superpotential is taken to be {\it an even function of the quark and lepton superfields} ($S_i$ and $T_j$), i.e. invariant under $R$-parity, 
to allow for $B$ and $L$ conservation and avoid automatically direct exchanges of squarks and sleptons between ordinary quarks and leptons.  

The bilinear terms $T_j^\dagger S_i$ are absent in minimal supersymmetric extensions of the SM.
They correspond, more generally, to a direct superpotential mass term for {\it vectorial quarks or leptons}, 
present only for left-handed and right-handed fields having the same gauge transformation properties.
It was considered because at that time the SM structure with quarks and leptons transforming as 
left-handed doublets and right-handed singlets was not confirmed yet, and Fayet wanted to allow for more general possibilities. 
Although absent in minimal SUSY extensions of the SM, such terms may be important in other situations. 
This is the case for the vectorial ``messenger'' 
quarks (and leptons) introduced to generate gluino masses from the messenger scale through radiative corrections \cite{glu},
as done now in the so-called ``GMSB" models.

Let us return to the superfields describing the usual SM quarks and leptons,
which are the left-handed doublets $S_i$ and right-handed singlets $\,T_j$.
The products $\,T_j^\dagger S_i$ are left-handed doublet superfields, as follows
\begin{equation}
\label{bilsup}
T_j^\dagger S_i\ \to\  \left\{\begin{array}{lll}
T^{\dagger}_{e_{a}}\,S_{a} &\hbox{for leptons, now written}
&\bar E_{a}\,L_{a}
\vspace{2mm}\\
T^{\dagger}_{d_{i}}\,S_{i}, \ \
T^{\dagger}_{u_{i}}\,S_{i}\ & \hbox{for quarks, now written}
&{\bar D_{i}}\,Q_{i},\ \
{\bar U_{i}}\,Q_{i}.
\end{array}\right.
\end{equation}

They are bilinear products of left-handed lepton ($S_{a}$) and quark ($S_{i}$) doublet 
superfields, with the conjugates of their right-handed superfield counterparts describing 
right-handed leptons ($T_a$) and quarks ($T_i$). These conjugates are left-handed singlet superfields 
now denoted as $T^{c}_{e_{a}}=\bar E_a, \, T^{c}_{d_{i}}=\bar D_i$ and $T^{c}_{u_{i}}=\bar U_i$, in modern notations.
These bilinear terms in (\ref{bilsup})
are then coupled in a supersymmetric and gauge invariant way 
to the two doublet Higgs superfields $\,S\,$ ($=H_1$) left-handed and $\,T$ right-handed (now replaced by $\,T^c=H_2$).
The corresponding $\,S\,T_j^\dagger S_i$ and $\,T^\dagger \,T_j^\dagger S_i\,$
trilinear tems in the superpotential (\ref{superpotterms}) read, for SM quarks and leptons,
\begin{equation}
\label{supot}
{\cal W}_{lq} \ =\  h_{e_{a}} \ S \ T^{\dagger}_{e_{a}}\,S_{a} \,+\, 
h_{d_{i}} \ S \ T^{\dagger}_{d_{i}}\,S_{i} \,+\,  
h_{u_{i}} \  T^{\dagger} \ T^{\dagger}_{u_{i}}\,S_{i} \ ,	
\label{earlypotential}	    
\end{equation}
i.e. in modern langage, as seen from (\ref{bilsup}),
\begin{equation}
{\cal W}_{lq} \ =\  h_{e_{a}} \ H_1\,\bar E_{a}\,L_{a}\,+\, h_{d_{i}} \ H_1\ {\bar D_{i}}\,Q_{i}\,-\,h_{u_{i}}
H_2\ {\bar U_{i}}\,Q_{i}\ \ .
\end{equation}

One should still add the $\,\mu\,H_1 H_2$ direct Higgs superpotential mass term 
when it is allowed by the symmetries considered. It should otherwise by replaced as in \cite{R} by a trilinear coupling 
with an extra singlet superfield, now written as $\,\lambda\,H_1H_2 S$, as in indicated in (\ref{superpot}).

The vacuum expectation values of the two Higgs doublets 
described by
$\,S\,$ and $\,T\,$ are
given (depending on how chiral superfields are normalized) by
\begin{equation}
\left\langle S \right\rangle \ =\ \left( \begin{array}{cc} v^{\prime \prime}/\sqrt 2 \vspace{.1truecm}\\ 0
\end{array} \right) \,,\ \
\left\langle T \right\rangle \ = \ \left( \begin{array}{cc} v^{\prime}/\sqrt 2 \vspace{.1truecm}\\ 0
\end{array} \right).
\end{equation}
They generate charged-lepton and down-quark masses, 
and up-quark masses,
given here by
$\,m_e\,=\,h_e\,v^{\prime\prime}/\sqrt 2\,,\ \,m_d\,=\,h_d\,v^{\prime\prime}/\sqrt 2\,,$ ~and 
$\,m_u\,=\,h_u\,v^{\prime}/\sqrt 2\,$,
~respectively (now $h_e\,v_1/\sqrt 2,\ h_d\,v_1/\sqrt 2$ ~and 
$\,h_u\,v_2/\sqrt 2\,$).

The correspondence between the earlier notations 
for doublet Higgs superfields and mixing angle, 
and modern ones, taken from the third reference in \cite{Fayet:2001xk},
is as follows:
\\
\begin{center}
\begin{tabular}{|ccc|} \hline 
&&\\ 
$
S=\left(\! \begin{array}{cc} S^0 \vspace{.1truecm}\\ S^-
\end{array}\! \right)\ \,\hbox{and}\ \ \,
T= \left(\! \begin{array}{cc} T^0 \vspace{.1truecm}\\ T^-
\end{array} \!\right)$	 &  $\!\!\!\longmapsto \!\!\!$   &  
$H_1 =\left( \!\begin{array}{cc} H_1^{\,0} \vspace{.1truecm}\\ H_1^{\,-}
\end{array} \!\right)\ \hbox{and}\  \,
H_2 =  \left(\! \begin{array}{cc} H_2^{\,+} \vspace{.1truecm}\\ H_2^{\,0}
\end{array} \!\right)$
                           \\  [.2 true cm] && \\
\  (left-handed) \ \  \ (right-handed) \  &   &   (both left-handed)   
         \\   [.2 true cm]   \hline  && \\ 
$ \hbox{\large{$\tan \,\delta$}} = \frac{<T^0>}{<S^0>} =
\frac{<\varphi'^0>}{<\varphi''^0>} = \displaystyle{v'\over v''}$ & 
$\!\!\! \longmapsto  \!\!\!$ &
$ \hbox{\large{$\tan \,\beta$}} = \frac{<H_2^{\,0}>}{<H_1^{\,0}>}\ = \
\frac{<h_2^{\,0}>}{<h_1^{\,0}>} =  \displaystyle{v_2\over v_1}
$
 \\ [.2 true cm] && \\ \hline
\end{tabular}
\end{center}

The whole construction showed that one could deal elegantly 
with spin-0 Higgs boson fields
(not a very popular ingredient at the time)
in the framework of spontaneously-broken supersymmetric theories.  

\begin{table}[t]
\caption{\ Minimal particle content of the MSSM taken from the first reference in \cite{Fayet:2001xk}.
\label{tab:SSM}}
\vspace{0.2cm}
\begin{center}
\begin{tabular}{|c|c|c|} \hline 
&&\\ [-0.2true cm]
Spin 1      
&\hbox{\hskip -1truecm Spin 1/2 \hskip -1truecm}    &Spin 0 \\ [.1 true cm]\hline 
&&\\ [-0.2true cm]
gluons  &\hbox{\hskip -1truecm gluinos ~$\tilde{g}$  \hskip -1truecm}      &\\
\,photon   \,     &\hbox{\hskip -1truecm photino ~$\tilde{\gamma}$ \hskip -1truecm } &\\ 
------------&$ - - - - - \, - $&-------------------- \\
 

$\begin{array}{c}
W^\pm\\ [.1 true cm]Z \\ 
\\ \\
\end{array} $

&
$\begin{array}{c}
\hbox {winos}  \ \widetilde W_{1,2}^{\,\pm} \\ 
[0 true cm]
\hbox {zinos } \ \ \widetilde Z_{1,2} \\ 
\\ 
\hbox {higgsino } \ \tilde h^0 
\end{array}$

& $\left.  \!\!\!\begin{array}{c}
H^\pm\\
[0 true cm] H\ \\
\\
h, \ A
\end{array}\!\!\right\} 
\begin{array}{c} \hbox {Higgs}\!\!\!\\ \hbox {bosons}\!\!\! \end{array}$  \\ &&\\ 
[-.1true cm]
\hline &&

\\ [-0.2cm]
&leptons  ~$l$   \hskip -1truecm    &sleptons  ~$\tilde l$ \\
&quarks ~$q$       &squarks   ~$\tilde q$\\ [-0.3 cm]&&
\\ \hline
\end{tabular}
\end{center}
\end{table}

The exact mass spectrum depends of course on the details 
of the supersymmetry breaking mechanism considered:
use of soft-breaking terms, possibly ``derived from supergravity", 
presence or absence of extra-$U(1)\,$ gauge fields 
and/or additional chiral superfields (as in the USSM or N/nMSSM), r\^ole of radiative corrections
involving messenger quarks, etc..
In any case, independently of the details of the
supersymmetry-breaking mechanism 
ultimately considered, 
we obtain the minimal particle content
of the Supersymmetric Standard Model, as given in 
\hbox{Table \ref{tab:SSM}}.

\subsection{Mass spectrum in the gauge and Higgs sector of the nMSSM}

More specifically the particle spectrum of the model described in \cite{R},
which corresponds in fact to the gauge and Higgs sector of the nMSSM
(with at this stage no or only a minimal breaking of the supersymmetry, limited to charged particles), is described below.
At first the gauge boson fields
\begin{eqnarray}
   A_{m}(x) &=&   \cos\theta\, V_{m}(x)
              - \sin\theta\,V^{3}_{m}(x) \,\ , 
              \nonumber \\
   Z_{m}(x)  &=&    \sin\theta\,V_{m}(x)
                  + \cos\theta \,V^{3}_{m}(x) \,\ , 
                  \nonumber \\ 
   W^{\pm}_{m}(x) &=& \frac{V^{1}_{m}(x) \mp i V^{2}_{m}(x)}{\sqrt{2}} \,\ ,
              \label{mass0}
\end{eqnarray}
where $A_{m}$ is the massless photon field. The masses of the heavy vector bosons are 
at lowest order
\begin{equation}
m_{Z}=\frac{1}{2} \sqrt{(g^{2}+g^{\prime 2})(v^{\prime 2}+v^{\prime \prime 2})}, \,\
m_{W}=\frac{g}{2} \sqrt{v^{\prime 2}+v^{\prime \prime 2}},
\label{mass1}
\end{equation}
where $\theta$ is related to the gauge boson mass ratio by
\begin{equation}
\cos \theta = \frac{m_{W}}{m_{Z}}.
\label{mass2}
\end{equation}

Their supersymmetric partners (gauginos) are defined by:
\begin{eqnarray}
   \lambda_{ \gamma}(x)   &=&     \cos\theta \,\lambda^{\prime}(x)
                   - \sin\theta \,\lambda^{3}_{A}(x) \,\ ,
                   \nonumber \\
    \lambda_{Z}(x)   &=&   \sin\theta \,\lambda^{\prime}(x)
                   + \cos\theta \,\lambda^{3}_{A}(x) \,\ ,
                   \nonumber \\
   \lambda^{\pm}(x) &=& \frac{\lambda^{1}_{A}(x) \mp i \lambda^{2}_{A}(x)}{\sqrt{2}} \,\ .
                    \label{mass3}
\end{eqnarray}
The photino $\lambda_{ \gamma}$, associated with the photon within a massless gauge multiplet
of supersymmetry,
is also here the massless Goldstone spinor from the spontaneous breaking of 
the supersymmetry. The latter role will of course disappear if soft terms breaking explicitly the supersymmetery are introduced, so as to make superpartners heavy.

The scalar-boson mass-eigenstate combinations are denoted by
\begin{eqnarray}
z&=&- \phi^{\prime}_{01} \sin \delta + \phi^{\prime \prime}_{01} \cos \delta , \nonumber \\
\omega &=& \frac{a- \imath b}{\sqrt{2}}, \nonumber \\
\phi &=& \phi^{\prime}_{01} \cos \delta + \phi^{\prime \prime}_{01} \sin \delta , \nonumber \\
w_{-}\!\!&=& \phi^{\prime}_{-} \cos \delta - \phi^{\prime \prime}_{-} \sin \delta . \nonumber \\
                    \label{mass4}
\end{eqnarray}
Their zeroth order masses are
\begin{eqnarray}
m_{z}&=&m_{Z}, \nonumber \\ 
m_{\omega}&=&m_{\phi}=\frac{h}{2} \sqrt{v^{\prime 2}+v^{\prime \prime 2}}, \nonumber \\
m_{w_{-}}\!\!&=&m_{W}.
\label{mass5}
\end{eqnarray}

The physical spinors are
\begin{eqnarray}
e_{-}&=&- \lambda^{-}_{L}+ \psi^{-}_{R}, \nonumber \\
E_{-}&=&- \lambda^{-}_{R}+ \psi^{-}_{L}, \nonumber \\
e_{0}&=& \imath p_{L}+ \left( \psi_{0} \cos \delta + \psi^{*}_{0} \sin \delta \right)_{R}, \nonumber \\
E_{0}&=& \left( \psi_{0} \cos \delta - \psi^{*}_{0} \sin \delta \right)_{L}- \lambda_{ZR}, \nonumber \\
\nu_{L}&=& \lambda_{\gamma L}.
\label{mass6}
\end{eqnarray}
Their masses are at lowest order
\begin{eqnarray}
m_{e_{-}}&=& \frac{gv^{\prime}}{\sqrt{2}}, \nonumber \\
m_{E_{-}}&=& \frac{gv^{\prime \prime}}{\sqrt{2}}, \nonumber \\
m_{e_{0}}&=&m_{\omega}=\frac{h}{2} \sqrt{v^{\prime 2}+v^{\prime \prime 2}}, \nonumber \\
m_{E_{0}}&=&\frac{1}{2} \sqrt{(g^{2}+g^{\prime 2})(v^{\prime 2}+v^{\prime \prime 2})}, \nonumber \\
m_{\nu}&=&0,
\label{mass7}
\end{eqnarray}
the latter vanishing exactly as a consequence of the continuous $R$-invariance, 
and of the role of this would-be ``neutrino'', i.e. in fact of the photino, as a Goldstone spinor.
The fields $E_{-},\ E_{0}$ and $e_{0}$ are associated with  heavy ``leptons''. $\nu_{L}$ is 
identified with the ``neutrino'' field, and $e_{-}$ with the ``electron'' field. 
The charged fermion masses are related to the mixing angle $\delta$ by
\begin{equation}
\tan \delta = \frac{m_{e_{-}}}{m_{E_{-}}}.
\label{mass8}
\end{equation}
There is here for the moment no mass splittings between bosons and fermions within neutral multiplets of supersymmetry.
The $Z$ remains at this stage mass-degenerate with a neutral Higgs boson (now called $h$),
 and the $W^\pm$ with a charged one (now called $H^\pm$).
The fermions of this initial model became the charginos and neutralinos of the supersymmetric extensions of the standard model \cite{ssm}, the mixing angle $\,\delta\,$ defined by $\tan \delta=v'/v"$ being now replaced by $\,\beta$, defined by the same formula  $\tan\beta=v_2/v_1$.

\section{Early searches for supersymmetry}
\label{sec:ess}

We present here the first studies on the experimental consequences of Supersymmetric 
Extensions of the Standard Model.

\subsection{\boldmath $R$-Hadrons}

SUSY's phenomenology started with the search for  "$R$-hadrons"~\cite{ff,ff2} in 1978, 
potentially the most easily detectable of the new particles associated with supersymmetry, 
as gluinos could have been massless or light, with their masses naturally protected from being large by the continous $R$-symmetry --~in contrast with squark and slepton masses, which could be much larger.

In this first phenomenological article Farrar and Fayet started defining "$R$-hadrons" in the following way \cite{ff}:
 ``The gluinos, 
which are flavor singlets and belong to a color $SU(3)$ octet, interact strongly with the octet of 
gluons and may combine with quarks, anti-quarks and gluons in much the same way as quarks presumably 
do, giving new hadronic states carrying one unit of $R$, which we call $R$-hadrons. Combined with 
$qqq$, $\bar{q}q$ or simply gluons, gluinos lead, respectively, to new bosonic states ($R$-baryons) 
and new fermionic states ($R$-mesons, $R$-glueballs) which are color-singlets with a flavor multiplet 
structure similar to ordinary baryons, mesons and glueballs.

These states should be produced in pairs in hadronic reactions. Naively, since gluinos-gluon 
interaction is like the quark-gluon interaction except for Clebsh-Gordan coefficients, one can 
expect $R$-hadron pair production comparable to the pair production of the corresponding 
hadrons of comparable mass".

The lightest of these $R$-odd hadrons is expected to be neutral and if gluinos are indeed massless, 
$1$-$1.5$ GeV or up to $2$ GeV is a reasonable estimate for the mass of the $R$-glueball, the production of 
light $R$-hadrons being naively expected at the millibarn level \cite{ff}.

In the second phenomenological article dealing with supersymmetry we find  \cite{ff2}: ``Those 
$R$-hadrons which are stable to strong and electromagnetic decays are nevertheless unstable. 
The decay into ordinary hadrons by emitting a new neutrino-like particle, the nuino (this may 
be the Goldstino, arising from the spontaneous breaking of the supersymmetry, or the photino, 
the fermionic partner of the photon). The corresponding lifetimes depend strongly on the class 
of models considered and, obviously, on the $R$-hadrons masses. We estimate \cite{ff}, in the 
simplest models, that $R$-hadrons in the interval $1-1.5$ GeV have lifetimes 
$\approx 10^{-12}-10^{-15}$ s or less". In this article they also evaluate  the cross section for the 
production of a pair of $R$-hadrons in $pN$ collisions, through the reaction $pN \rightarrow R \bar{R}$. 
They get, for a $R$-hadron mass of 2 GeV
\begin{equation}
\sigma_{pN \rightarrow \bar{R}R+X}\leq 40 \ \mu b \,\ \mbox{at}\  \sqrt{s}=27\  \mbox{GeV}.
\end{equation}
Their properties and production rates are model-dependent, partly because 
the understanding of hadrons was not yet very good at the time.

In the late 1970s, several fixed-target experiments obtained upper limits on the pair-production cross 
section of hadrons decaying to a final state with missing energy-momentum \footnote{This 
production was studied in \cite{ff,ff2}.}, excluding a range of $m_{\tilde{g}}$ (of typically less than a few GeV's),
depending 
on the lifetime. All experiments give negative results (for more details see \cite{tata}), and we know today 
that gluinos should be quite heavy ... 

\subsection{Pair production of spin-0 leptons in \boldmath $e^{+}e^{-}$ annihilations.}

The pair-production of spin-0 leptons 
directly in $e^{+}e^{-}$ annihilations at the SPEAR collider was studied in 1979 \cite{ff3}, and this process was used soon after to search for selectrons and smuons at PETRA.
This article concentrated on the spin-$0$ partners of the electron and muon. These channels were chosen because their electromagnetic properties are completely determined by supersymmetry, production cross sections are substantial (above threshold) and they have a very distinctive signal. 

The authors  also considered in 
their study the signal in the experiment: ``Spin-0 leptons are unstable and decay extremely 
quickly into the corresponding lepton by emission of a photino or goldstino". The corresponding reaction is
shown in Fig.~\ref{fig:e-e+}:
\begin{eqnarray}
e^{+}e^{-} &\rightarrow& \mbox{Pair of sleptons} \nonumber \\
&\rightarrow& \mbox{Non-coplanar pair (}e^{+}e^{-}, \,\ \mu^{+}\mu^{-} \,\ \mbox{or}\,\ \tau^{+}\tau^{-} \mbox{)} \nonumber \\
&&+ \mbox{\ 2 unobserved photinos or goldstinos}.
\end{eqnarray}
Following \cite{ff3}, ``This would lead to non-coplanar events with half of the energy missing on average, 
since the photinos or goldstinos would be unobserved".
The cross section for selectrons ($K\geq 1$) or smuons ($K=0$) is  given by:
\begin{equation}
\frac{d \sigma (e^{-}e^{+} \rightarrow s \bar{s}+t \bar{t})}{d( \cos \theta )}=\frac{\pi \alpha^{2}\beta^{3}\sin^{2} \theta}{4s} 
\left[ 1+ \left( 1- \frac{4K}{1-2 \beta \cos \theta + \beta^{2}}
\right)^{2} \right]\,,
\end{equation}
the second term being associated with photino exchanges in the $t$ channel, for the pair-production of selectrons.
Staus were not considered, as ``For spin-0 $\tau$'s, the limit is inevitably lower, since non-coplanar 
$\tau^{+}\tau^{-}$ pairs are not observed directly, but by looking at $\mu^{\pm}+$ 
hadrons + (missing energy) events".

The production of such no-coplanar $e^{+}e^{-}$ (or, similarly,  $\mu^{+}\mu^{-}$) pairs with missing energy-momentum 
is not larger than $\sim 10$ pb 
at  $E_{b}\approx 3.6$ GeV (the available beam energy at the time), leading to the conclusion that spin-0 electrons and muons had to be heavier than $\,\sim 3\,\frac{1}{2}\,$ GeV.
These lower limits soon increased up to about 15 GeV
at PETRA, and are now larger than 90 GeV, from LEP experiments ...

For spin-0 quarks which could be pair-produced in $e^+ e^-$ annihilations,
in addition to ordinary to spin-$\frac{1}{2}$ ones, on would get altogether
the ratio
\begin{equation}
R= \frac{\sigma(e^{+}e^{-} \rightarrow \mbox{hadrons})}{\sigma(e^{+}e^{-} \rightarrow \mu^{+}\mu^{-})}=
\frac{3}{2}\ \sum_{i}q^{2}_{i}= \frac{11}{2},
\end{equation}
$q_{i}$ being the charges of the $u,d,c,s$ and $b$ quarks.

\section{Conclusion}
\label{sec:concl}
We reviewed the problems one had to face for the construction of a supersymmetric 
extension of the SM, in the early days of the supersymmetry theory. We presented
the first supersymmetric extension of the standard model constructed by P. Fayet in the years 1974-77, and, as  
a last topic, the first studies of experimental consequences of supersymmetry following from this work. 
We hope that this analysis will be useful to those interested in supersymmetric 
extensions of the standard model.

\vbox{
\begin{center}
{\bf Acknowledgments} 
\end{center}
This work was partially financed by the Brazilian funding agency
CNPq, under contract number 
309564/2006-9. We are grateful to Pierre Fayet for sending us so many interesting informations about the early days of 
the Supersymmetric extensions of the Standard Model, where I havew learnt a lot. Without his help, I could not have written 
this review.
}

\appendix

\section{Lagrangian density}
\label{apen:lagrangian}

The Lagrangian density coupling chiral to gauge superfields, invariant under local gauge transformations, is given by:
\begin{eqnarray}
&&\int d^{4}x \int d^{2} \theta \bar{ \Phi} e^{gV_{WZ}} \Phi= \int d^{4}x \left \{ 
-({\cal D}_{m}A)^{\dagger}({\cal D}^{m}A) \right. \nonumber \\
&-& \left. \imath \bar{\psi}\bar{\sigma}^{m}{\cal D}_{m}\psi +
\bar{F}F- \imath g \sqrt{2}T^{a}(A \bar{ \lambda}^{a} \bar{ \psi}- \bar{A} \lambda^{a} \psi)+ 
gT^{a}D^{a}\bar{A}A
\right \} \,\ . \nonumber \\ 
\label{d1}
\end{eqnarray}
where
\begin{eqnarray}
{\cal D}_{m}A&=& \partial_{m}A+ \imath gT^{a}v^{a}_{m}A \,\ , \nonumber \\
{\cal D}_{m} \psi&=& \partial_{m}\psi + \imath gT^{a}v^{a}_{m}\psi \,\ . \nonumber \\
\end{eqnarray}
The superpotential terms are obtained as
\begin{eqnarray}
&& \int d^{2} \theta \left( \lambda_{i} \Phi_{i}+ \frac{m_{ij}}{2} m \Phi_{i} \Phi_{j} + 
\frac{g_{ijk}}{3} \Phi_{i} \Phi_{j} \Phi_{k} \right) = \bar{A}_{i} \Box A_{i}+ \bar{F}_{i}F_{i}+i \partial_{m} \bar{ \psi}_{i} 
\bar{ \sigma}^{m} \psi_{i}+
\lambda_{i}F_{i} \nonumber \\ &+&m_{ij}A_{i}F_{j} 
-\frac{m_{ij}}{2} \psi_{i} \psi_{j}+
g_{ijk} \left( F_{i}A_{j}A_{k}- \psi_{i} \psi_{j} A_{k} \right)  \,\ . \nonumber \\
\label{aq2} 
\end{eqnarray}
The supersymmetric Yang-Mills Lagrangian density is given by
\begin{eqnarray}
\frac{1}{4k} \int d^{2} \theta Tr(W^{\alpha}W_{\alpha})+ \frac{1}{4k} \int d^{2} \bar{\theta} 
+ Tr(\bar{W}_{ \dot{ \alpha}} \bar{W}^{ \dot{ \alpha}})= \left[ -\frac{1}{4} F^{a}_{mn}F^{a mn}
- \imath \lambda^{a} \sigma^m {\cal D}_{m} \bar{ \lambda}^{a} + \frac{1}{2} D^{a}D^{a} \right] \,\ , \nonumber \\
\label{d3333}
\end{eqnarray}
where
\begin{eqnarray}
F^{a}_{mn}&=& \partial_{m}v^{a}_{n}- \partial_{n}v^{a}_{m}-gt^{abc}v_{m}^{b}v_{n}^{c}, \nonumber \\
{\cal D}_{m} \lambda^{a}&=& \partial_{m} \lambda^{a}-gf^{abc}A^{b}_{m} \lambda^{c}.
\end{eqnarray}

\section{\boldmath $R$ Symmetry}
\label{apen:rsymmetry}

$R$-symmetry is better understood with the superspace
formalism. It  is a continuous $U(1)$ 
symmetry acting on the supersymmetry generator, parame\-trized by $\alpha$. The corresponding
operator will be denoted as ${\bf R}$. $R$-symmetry 
acts on the superspace coordinate $\theta$, $\bar{\theta}$ as
follows \cite{wb} 
\begin{eqnarray}  
{\bf R} \theta &=& e^{-i \alpha} \theta, \nonumber \\ 
{\bf R} \bar{\theta} &=& e^{i \alpha} \bar{\theta}. 
\label{The R-Invariance prop 1} 
\end{eqnarray} 
$\theta$ has $R$-charge $\mathrm{R}( \theta )= -1$, while
$\bar{\theta}$ has $\mathrm{R}( \bar{\theta} )=1$.

The operator ${\bf R}$ acts on left-handed chiral superfields
$\Phi(x,\theta,\bar{\theta})$ and (right-handed) 
anti-chiral ones
$\bar{\Phi}(x,\theta,\bar{\theta})$ in the
following way~\cite{susy,wb} 
\begin{eqnarray}
   {\bf R} \Phi(x,\theta,\bar{\theta}){\bf R}^{-1}  &=& e^{i n_{\Phi}\alpha}
\Phi(x, e^{-i\alpha}\theta ,e^{i\alpha}\bar{\theta} ),\label{eq3}\\
   {\bf R} \bar{ \Phi}(x,\theta,\bar{\theta}){\bf R}^{-1}  &=&
          e^{- i n_{\Phi}\alpha}\bar{ \Phi}(x, e^{-i\alpha}\theta ,
          e^{i\alpha}\bar{\theta} ),
          \label{The R-Invariance prop 2}
\end{eqnarray} 
where $n_{ \Phi}$ is the $R$-charge of the chiral
superfield. 
It acts on vectorial (gauge) superfields  by
\begin{eqnarray}
   {\bf R} V(x,\theta,\bar{\theta}){\bf R}^{-1} &=&
V(x, e^{-i\alpha}\theta , e^{i\alpha}\bar{\theta} ).
         \label{The R-Invariance prop 3}
\end{eqnarray}

The expansions of the superfields in terms of  $\theta$ and
$\bar{\theta}$, see \cite{wb}, are given by 
\begin{eqnarray} 
\Phi(x, \theta, \bar{ \theta}) &=& A(x) + \sqrt{2} \theta \psi (x) + \theta \theta
F(x) \nonumber \\ 
& & + i  \theta \sigma^m \bar{ \theta} \partial_m A(x) -
\frac{i}{ \sqrt{2}}( \theta \theta ) \partial_m \psi (x) \sigma^m
\bar{ \theta} \nonumber \\
& &+ \frac{1}{4} (\theta \theta)( \bar{ \theta}
\bar{ \theta} ) \Box A(x),\label{eq1} 
\\ [2mm]
\bar{ \Phi}(x, \theta, \bar{ \theta}) &=& \bar{A}(x) + \sqrt{2} \bar{ \theta} \bar{ \psi}
(x) + \bar{ \theta} \bar{ \theta} \bar{F}(x) \nonumber \\ 
& &-i \theta
\sigma^m \bar{ \theta} \partial_m \bar{A}(x) + \frac{i}{
\sqrt{2}}( \bar{ \theta} \bar{ \theta} ) \theta \sigma^m
\partial_m \bar{ \psi} (x) \nonumber \\ 
& &+ \frac{1}{4} (\theta \theta)(
\bar{ \theta} \bar{ \theta} ) \Box \bar{A}(x),\label{eq2}
\\ [3mm]
V_{WZ}(x, \theta, \bar{ \theta}) &=& -\theta \sigma^{m} \bar{
\theta} A_{m}(x) + i( \theta \theta ) \bar{ \theta} \bar{
\lambda}(x) - i(\bar{ \theta} \bar{ \theta} )\theta \lambda(x) \nonumber \\ 
& &+ \frac{1}{2} (\theta \theta )( \bar{ \theta} \bar{
\theta} )D(x)\ \ \ \ \mbox{\small (in the Wess-Zumino gauge)}.
\label{exp1}
\end{eqnarray}
$A(x),F(x)$ and $D(x)$ are scalar fields;
$\psi(x)$ and $\lambda(x)$ are fermion fields, while $A_{m}(x)$ is
 vector field.

Combining (\ref{eq3}) and (\ref{eq1}) we get the
transformations for the field components:
\begin{eqnarray} \left.
\begin{array}{lcr}
A(x)    &\stackrel{{\bf R}}{\longmapsto}&       e^{in_{\Phi}\alpha} A(x) \\
\psi (x) &\stackrel{{\bf R}}{\longmapsto}&e^{i
\left( n_{\Phi}-1 \right) \alpha}\psi (x) \\
F(x)    &\stackrel{{\bf R}}{\longmapsto}&       e^{i \left( n_{\Phi}-1
\right)\alpha} F(x)
             \end{array} \right\}.
           \label{The R-Invariance prop 4a}
\end{eqnarray} 
 From (\ref{The R-Invariance prop 3}) and (\ref{exp1}), the  field
components in the vector superfield transform as 
\begin{eqnarray}
   \left.  \begin{array}{lcr}
A_{m}(x) &\stackrel{{\bf R}}{\longmapsto}&       A_{m}(x) \\
\lambda (x)     &\stackrel{{\bf R}}{\longmapsto}&
    e^{i\alpha} \lambda (x) \\
\bar{\lambda}(x)     &\stackrel{{\bf R}}{\longmapsto}&
    e^{-i\alpha} \bar{\lambda}(x) \\
D(x)       &\stackrel{{\bf R}}{\longmapsto}&       D(x)
          \end{array} \right\}.
          \label{The R-Invariance prop 5}
\end{eqnarray} 
These transformations
may be rewritten  in terms of 4-components spinors as \cite{susy,moreau} 
\begin{equation}
\begin{array}{rcl}
A_{m}(x) &\stackrel{{\bf R}}{\longmapsto}  & A_{m}(x), \\
\Lambda(x) &\stackrel{{\bf R}}{\longmapsto}  & e^{i \gamma_5 \alpha}
\ \Lambda(x), \\
D(x) &\stackrel{{\bf R}}{\longmapsto}  & D(x),
\end{array}
\begin{array}{rcl}
A(x) &\stackrel{{\bf R}}{\longmapsto}  & e^{in_{\Phi} \alpha} \ A(x), \\
\bar{A}(x) &\stackrel{{\bf R}}{\longmapsto}  &
e^{-in_{\Phi} \alpha} \ \bar{A}(x), \\
\Psi(x) &\stackrel{{\bf R}}{\longmapsto}  & e^{i \gamma_5
(n_{\Phi}-1)\alpha} \ \Psi(x), \\
F(x) &\stackrel{{\bf R}}{\longmapsto}  & e^{i(n_{\Phi}-2) \alpha} \ F(x), \\
\bar{F}(x) &\stackrel{{\bf R}}{\longmapsto}  &
e^{-i(n_{\Phi}-2) \alpha} \ \bar{F}(x).
\end{array}
\label{Rsym4} 
\end{equation}
The Majorana spinor $\Lambda$
represents gauginos, and $\Psi(x)$ the Dirac
spinors for quarks and leptons.

For products of left-handed chiral superfields, 
\begin{eqnarray}
  {\bf R} \prod_{a}\;\Phi_{a}(x,\theta ,\bar{\theta})
    &=& e^{i\sum_{a}n_{a}\alpha}\,\prod_{a}\Phi_{a}
    (x, e^{-i\alpha}\theta ,e^{i\alpha}\bar{\theta} ).
      \nonumber \\
\end{eqnarray}  

\section{Feynman Rules for SQCD}
\label{apen:frsqcd}

 We derive here the Feynman Rules for SQCD presented in this article.

\subsection{Interaction from ${\cal L}_{gauge}$}

We can rewrite ${\cal L}_{gauge}$, see \cite{dress,tata} and Eq.(\ref{d3333}), in the following way
\begin{eqnarray}
{\cal L}_{gauge}&=& {\cal L}_{cin}+{\cal L}_{gaugino}+{\cal L}^{gauge}_{D}.
\end{eqnarray}
where
\begin{eqnarray}
{\cal L}_{cin}&=&{\cal L}^{SU(3)}_{cin}+{\cal L}^{SU(2)}_{cin}+{\cal L}^{U(1)}_{cin}, \nonumber \\
{\cal L}_{gaugino}&=&{\cal L}^{SU(3)}_{gaugino}+{\cal L}^{SU(2)}_{gaugino}+{\cal L}^{U(1)}_{gaugino}, \nonumber \\
 {\cal L}^{gauge}_{D}&=&{\cal L}^{SU(3)}_{D}+{\cal L}^{SU(2)}_{D}+{\cal L}^{U(1)}_{D},
\end{eqnarray}
the first part is given by:
\begin{equation}
{\cal L}^{SU(3)}_{cin}=- \frac{1}{4}G^{a}_{mn}G^{amn},
\label{usualqcd}
\end{equation}
with
\begin{equation}
G^{a}_{mn}= \partial_{m}g^{a}_{n}-\partial_{n}g^{a}_{m}-g_{s}f^{abc}g^{b}_{m}
g^{c}_{n}, 
\label{usualqcd1}
\end{equation}
$g_{s}$ is the strong coupling constant and $f^{abc}$ are the totally antisymmetric structure constants of 
$SU(3)$. The second term can be rewritten as
\begin{equation}
{\cal L}^{SU(3)}_{gaugino}=- \imath \overline{\lambda^{a}_{C}}\bar{\sigma}^{m}{\cal D}_{m}\lambda^{a}_{C},
\label{gaugino1}
\end{equation}
where
\begin{equation}
{\cal D}_{m}\lambda^{a}_{C}= \partial_{m}\lambda^{a}_{C}-g_{s}f^{abc}\lambda^{a}_{C}g^{c}_{m}.
\label{gaugino2}
\end{equation}
The last term is given by
\begin{equation}
{\cal L}^{SU(3)}_{D}= \frac{1}{2}D^{a}_{C}D^{a}_{C}.
\end{equation}

\subsubsection{Gluon self-interactions}

These interactions are derived from Eqs.(\ref {usualqcd},\ref{usualqcd1}), the same as in usual QCD, leading to
the same Feynman rules.

\subsubsection{Gluino-gluino-gluon interactions}

This interaction is obtained  from Eqs.(\ref{gaugino1}, \ref{gaugino2}), combining into
\begin{equation}
{\cal L}_{gaugino}={\cal L}_{cin}+{\cal L}_{\tilde{g}\tilde{g}g},
\end{equation}
where
\begin{eqnarray}
{\cal L}_{cin}&=& \imath ( \partial_{m} \overline{\lambda^{a}}_{C}) \bar{\sigma}^{m}\lambda^{a}_{C}, \nonumber \\ 
{\cal L}_{\tilde{g}\tilde{g}g}&=& \imath g_{s}f^{abc}  \overline{\lambda^{a}}_{C} \bar{\sigma}^{m}\lambda^{b}_{C}g^{c}_{m},
\end{eqnarray}
the first term gives the kinetic term for gluinos, and the last one the gluino-gluino-gluon interaction.

Considerating the four-component Majorana spinor for the gluino, 
\begin{equation}
\Psi(\tilde{g}^{a}) \;=\; \left( \begin{array}{r} - \imath \lambda^{a}_{C}(x) \\
                    \imath \overline{\lambda^{a}}_{C}(x) \end{array} \right) \,\ ,
                        \label{gluino spinor}
\end{equation}
we can rewrite 
${\cal L}_{\tilde{g}\tilde{g}g}$ in the following way
\begin{equation}
{\cal L}_{\tilde{g}\tilde{g}g}= \frac{\imath}{2} g_{s}f^{bac} \bar{\Psi}(\tilde{g}^{a})\gamma^{m}\Psi(\tilde{g}^{b})g^{c}_{m}.
\label{regrafeynman1}
\end{equation} 
Owing to the Majorana nature of the gluino one must multiply 
by 2 to obtain the Feynman rule (or add the graph with 
$\tilde{g} \leftrightarrow \bar{\tilde{g}}$)!

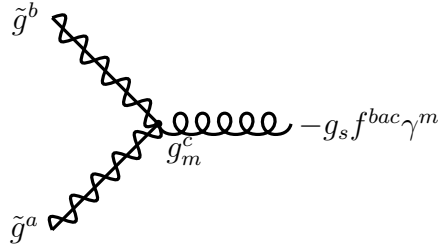
\begin{figure}[h]
\begin{center}
\begin{picture}(150,100)(0,0)
\SetWidth{1.2}
\Line(10,10)(50,50)
\Photon(10,10)(50,50){4}{5}
\Text(5,10)[r]{$\tilde{g}^{a}$}
\Line(10,90)(50,50)
\Photon(10,90)(50,50){4}{5}
\Text(5,90)[r]{$\tilde{g}^{b}$}
\Vertex(50,50){1.5}
\Text(60,40)[]{$g^{c}_{m}$}
\Gluon(100,50)(50,50){4}{5}
\Text(130,50)[]{$-g_{s}f^{bac}\gamma^{m}$}
\end{picture}
\end{center}
\caption{Feynman rule for the vertex gluino-gluino-gluon.}
\label{fig1}
\end{figure}

The above equation induces the Feynman rule for the vertex gluino-gluino-gluon given 
in Fig. \ref{fig1}, as obtained in \cite{dress,tata}.

\subsection{Interaction from ${\cal L}_{\mbox{\small \bf quarks}}$}

The interactions within the strong sector are obtained from the following Lagrangian densities
\begin{eqnarray}
{\cal L}_{qqg}&=&g_{s} \bar{Q} \bar{\sigma}_{m}T^{a}Qg^{am}+
g_{s} \overline{u^{c}} \bar{\sigma}_{m}\bar{T}^{a}u^{c}g^{am}+
g_{s} \overline{d^{c}} \bar{\sigma}_{m}\bar{T}^{a}d^{c}g^{am} \nonumber \\
{\cal L}_{ \tilde{q} \tilde{q}g}&=&\imath g_{s} \left( 
\bar{\tilde{Q}}T^{a}\partial_{m}\tilde{Q}-
\tilde{Q}T^{a}\partial_{m}\bar{\tilde{Q}} \right) g^{am}+
\imath g_{s} \left( 
\overline{\tilde{u}^{c}}\bar{T}^{a}\partial_{m}\tilde{u}^{c}-
\tilde{u}^{c}\bar{T}^{a}\partial_{m}\overline{\tilde{u}^{c}} \right) g^{am}
\nonumber \\ &+&
\imath g_{s} \left( 
\overline{\tilde{d}^{c}}\bar{T}^{a}\partial_{m}\tilde{d}^{c}-
\tilde{d}^{c}\bar{T}^{a}\partial_{m}\overline{\tilde{d}^{c}} \right) g^{am}, \nonumber \\
{\cal L}_{q \tilde{q} \tilde{g}}&=&- \imath \sqrt{2}g_{s} \left(
\bar{Q}T^{a}\tilde{Q}\overline{\lambda^{a}_{C}}-
\bar{\tilde{Q}}T^{a}Q\lambda^{a}_{C} \right) -
\imath \sqrt{2}g_{s} \left(
\overline{u^{c}}\bar{T}^{a}\tilde{u}^{c}\overline{\lambda^{a}_{C}}-
\overline{\tilde{u}^{c}}\bar{T}^{a}u^{c}\lambda^{a}_{C} \right) 
\nonumber \\
 &-&
\imath \sqrt{2}g_{s} \left(
\overline{d^{c}}\bar{T}^{a}\tilde{d}^{c}\overline{\lambda^{a}_{C}}-
\overline{\tilde{d}^{c}}\bar{T}^{a}d^{c}\lambda^{a}_{C} \right), 
\nonumber \\
{\cal L}_{ \tilde{q} \tilde{q}gg}&=&-g_{s}^{2}\bar{\tilde{Q}}T^{a}T^{b}\tilde{Q}g^{a}_{m}g^{bm}
-g_{s}^{2}\overline{\tilde{u}^{c}}\bar{T}^{a}\bar{T}^{b}\tilde{u}^{c}g^{a}_{m}g^{bm}-
g_{s}^{2}\overline{\tilde{d}^{c}}\bar{T}^{a}\bar{T}^{b}\tilde{d}^{c}g^{a}_{m}g^{bm}.
\label{quarksint}
\end{eqnarray}
Where $T^{a}_{rs}$ are the color triplet generators, one must use 
\begin{equation}
\bar{T}^{a}_{rs}=- T^{* a}_{rs}=-T^{a}_{sr},
\label{antitripletrepresentation}
\end{equation}
for the color anti-triplet generators.

\subsubsection{Quark--quark--gluon interaction}

\begin{figure}[h]
\begin{center}
\begin{picture}(150,100)(0,0)
\SetWidth{1.2}
\Line(10,10)(50,50)
\Text(5,10)[r]{$q_{r}$}
\Line(10,90)(50,50)
\Text(5,90)[r]{$q_{s}$}
\Vertex(50,50){1.5}
\Text(60,40)[]{$g^{a}_{m}$}
\Gluon(100,50)(50,50){4}{5}
\Text(130,50)[]{$- \imath g_{s} T^{a}_{rs} \gamma^{m}$}
\end{picture}
\end{center}
\caption{Feynman rule for the vertice quark-quark-gluon in the MSSM.}
\label{fig2}
\end{figure}
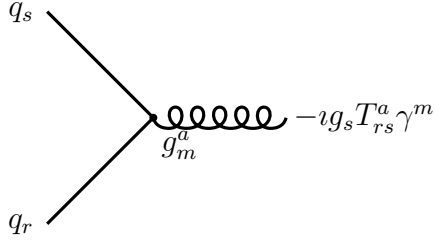

This interaction comes from the first Lagrangian density in Eq.(\ref{quarksint}),  and may be rewritten as
\begin{equation}
{\cal L}_{qqg}=g_{s}( \bar{u}_{r}\bar{\sigma}^{m}T^{a}_{rs}u_{s}+  \bar{d}_{r}\bar{\sigma}^{m}T^{a}_{rs}d_{s}+
\overline{u^{c}}_{r}\bar{\sigma}^{m}\bar{T}^{a}_{rs}u^{c}_{s}+  \overline{d^{c}}_{r}\bar{\sigma}^{m}\bar{T}^{a}_{rs}d^{c}_{s})g^{a}_{m},
\label{qqgmssm}
\end{equation}
$u$ and $d$ are color triplets while $u^{c}$ and $d^{c}$ are anti-triplets. $r$ and $s$ are color indices.  Using 
Eq.(\ref{antitripletrepresentation}) we can rewrite (\ref{qqgmssm}) as
\begin{equation}
{\cal L}_{qqg}=g_{s}( \bar{u}_{r}\bar{\sigma}^{m}T^{a}_{rs}u_{s}+  \bar{d}_{r}\bar{\sigma}^{m}T^{a}_{rs}d_{s}+
u^{c}_{r}\sigma^{m}T^{a}_{rs}\overline{u^{c}}_{s}+d^{c}_{r}\sigma^{m}T^{a}_{rs}\overline{d^{c}}_{s})g^{a}_{m}. 
\end{equation}
With the four-component quark Dirac spinor ($q=u,d$) given by 
\begin{equation}
\Psi(q) \;=\; \left( \begin{array}{r} q_{L}(x) \\
                    \overline{q^{c}}_{L}(x) \end{array} \right) \,\ , 
                        \label{quarks spinors}
\end{equation}
so that
\begin{equation}
{\cal L}_{qqg}=-g_{s}\sum_{q=u,d} \bar{\Psi}(q_{r})\gamma^{m}T^{a}_{rs}\Psi(q_{s})g^{a}_{m},
\label{regrafeynman2}
\end{equation}
we obtain the Feynman rule for the vertex $qqg$ in Fig. \ref{fig2}, as given in \cite{dress,tata}.

The gluino-gluon interaction is similar the quark-gluon interaction, as seen from Figs. \ref{fig1} and  \ref{fig2},
which leads to the idea of ``$R$-hadrons'' discussed in \cite{ff,ff2}.

\subsubsection{Squark--squark--gluon interaction}

This interaction is obtained from the term in the second line of Eq.(\ref{quarksint}), rewritten as
\begin{eqnarray}
{\cal L}_{\tilde{q}\tilde{q}g}&=& \imath g_{s}(  \bar{\tilde{u}}_{r}T^{a}_{rs}(\partial_{m}\tilde{u}_{s})-
(\partial_{m}\overline{\tilde{u}}_{r}) T^{a}_{rs}\tilde{u}_{s}+  \bar{\tilde{d}}_{r}T^{a}_{rs}(\partial_{m}\tilde{d}_{s})-
(\partial_{m}\overline{\tilde{d}}_{r}) T^{a}_{rs}\tilde{d}_{s} \nonumber \\ &+&
\overline{\tilde{u}^{c}}_{r}\bar{T}^{a}_{rs}(\partial_{m}\tilde{u}^{c}_{s})-
(\partial_{m}\overline{\tilde{u}^{c}}_{r}) \bar{T}^{a}_{rs}\tilde{u}^{c}_{s}+  
\overline{\tilde{d}^{c}}_{r}\bar{T}^{a}_{rs}(\partial_{m}\tilde{d}^{c}_{s})-
(\partial_{m}\overline{\tilde{d}^{c}}_{r}) \bar{T}^{a}_{rs}\tilde{d}^{c}_{s})g^{am}. \nonumber \\
\label{sqsqgmssm}
\end{eqnarray}
Using 
$\ 
  A\ {\!\stackrel{\leftrightarrow}{\partial}_{m}\!} B = A\,(\partial_{m} B) - (\partial_{m} A)\,B \,$,
we can rewrite our Lagrangian density in the following way
\begin{equation}
{\cal L}_{\tilde{q}\tilde{q}g}= \imath g_{s} \sum_{q=u,d} (
\tilde{q}_{Lr}^{*}T^{a}_{rs}\,  {\!\stackrel{\leftrightarrow}{\partial}_{m}\!} \, \tilde{q}_{Ls}-
\tilde{q}_{Rr}^{*}T^{a}_{rs}\,  {\!\stackrel{\leftrightarrow}{\partial}_{m}\!}\,  \tilde{q}_{Rs} )g^{am}\,.
\end{equation}
The relative minus sign between the $\tilde{q}_{L}$ and $\tilde{q}_{R}$ terms
is due to the fact that $\tilde{q}_{R}$ are colour antitriplets with
colour generator given in  (\ref{antitripletrepresentation}). We use a similar notation 
as in \cite{tata}, with $\tilde{q}^{*}$ creating the scalar quark $\tilde{q}$, while $\tilde{q}$ 
destroys the scalar quark $\tilde{q}$.

\begin{figure}[h]
\begin{center}
\begin{picture}(150,100)(0,0)
\SetWidth{1.2}
\DashLine(10,10)(50,50){4.5}
\Text(5,10)[r]{$\tilde{q}_{r}$}
\DashLine(50,50)(10,90){4.5}
\Text(5,90)[r]{$\tilde{q}_{s}$}
\Vertex(50,50){1.5}
\Text(60,40)[]{$g^{a}_{m}$}
\Gluon(100,50)(50,50){4}{5}
\Text(150,50)[]{$- \imath g_{s}T^{a}_{rs}(k_i^{} + k_j^{})^{m}$}
\end{picture}
\end{center}
\caption{Feynman rule for the vertice squark-squark-gluon in the MSSM.}
\label{fig3}
\end{figure}
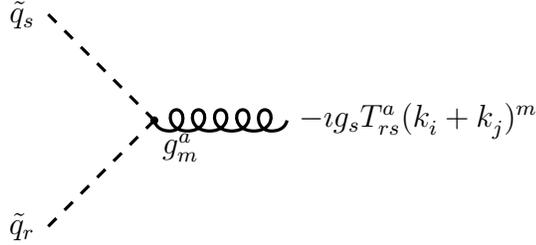

Generalization to six quark flavors, we get
\begin{equation}
{\cal L}_{\tilde{q}\tilde{q}g}= \imath g_{s} \sum_{q=u,d} \sum_{p=1}^{6} (
\tilde{q}_{Lpr}^{*}T^{a}_{rs}\,  {\!\stackrel{\leftrightarrow}{\partial}_{m}\!} \, \tilde{q}_{Lps}-
\tilde{q}_{Rpr}^{*}T^{a}_{rs}\,  {\!\stackrel{\leftrightarrow}{\partial}_{m}\!}\,  \tilde{q}_{Rps} )g^{am}
\label{squarksflavor}
\end{equation}
The corresponding Feynman rule are obtained from
\begin{equation}
  \tilde{q}_{j}^{*}\,  {\!\stackrel{\leftrightarrow}{\partial}^{m}\!}\,  \tilde{q}_{i} = \imath \,\ (k_i^{} + k_j^{})^{m}
\label{regrafeynman3}
\end{equation}
where $k_i^{}$ and $k_j^{}$ are the four--momenta of $\tilde{q}_{i}$ and $\tilde{q}_{j}$
in direction of the charge flow. This gives the  Feynman rule in Fig. \ref{fig3}, 
in agreement with \cite{dress,tata}.

\subsubsection{Squark-squark-gluon-gluon interaction}
\begin{figure}[h]
\begin{center}
\begin{picture}(150,100)(0,0)
\SetWidth{1.2}
\DashLine(10,10)(50,50){4.5}
\Text(5,10)[r]{$\tilde{q}_{r}$}
\DashLine(10,90)(50,50){4.5}
\Text(5,90)[r]{$\tilde{q}_{t}$}
\Vertex(50,50){1.5}
\Gluon(90,90)(50,50){4}{5}
\Text(100,90)[b]{$g^{a}_{m}$}
\Gluon(90,10)(50,50){4}{5}
\Text(100,10)[b]{$g^{b}_{n}$}
\Text(130,50)[]{$- \imath g^{2}_{s} \left( \frac{\delta_{ab}\delta_{rt}}{3}+d^{abc}T^{c}_{rt} \right)g_{mn}$}
\end{picture}
\end{center}
\caption{Feynman rule for the squark-squark-gluon-gluon vertex in the  MSSM.}
\label{fig4}
\end{figure}
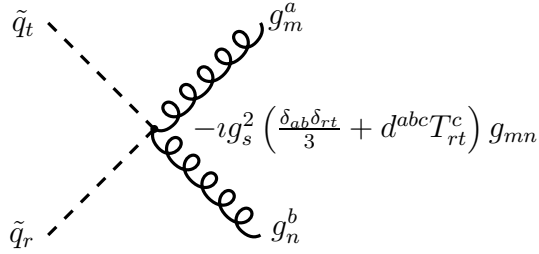

This interaction comes from the last line in  Eq.(\ref{quarksint}), which can be written as
\begin{eqnarray}
{\cal L}_{\tilde{q}\tilde{q}gg}&=&-g^{2}_{s}(
\bar{\tilde{u}}_{r}T^{a}_{rs}T^{b}_{st}\tilde{u}_{t}+
\bar{\tilde{d}}_{r}T^{a}_{rs}T^{b}_{st}\tilde{d}_{t}+
\overline{\tilde{u}^{c}}_{r}\bar{T}^{a}_{rs}\bar{T}^{b}_{st}\tilde{u}^{c}_{t}+
\overline{\tilde{d}^{c}}_{r}\bar{T}^{a}_{rs}\bar{T}^{b}_{st}\tilde{d}^{c}_{t})g^{a}_{m}g^{bm}. \nonumber \\
\label{sgsgggmssm}
\end{eqnarray}
Using the formula for $SU(3)$ generators
\begin{equation}
T^{a}_{rs}T^{b}_{st}= \frac{1}{6}\delta_{ab}\delta_{rt}+ \frac{1}{2}(d^{abc}+ \imath f^{abc})T^{c}_{rt},
\end{equation}
we can rewrite our Lagrangian density in the following way (including the generalization to six flavors as in (\ref{squarksflavor})):
\begin{eqnarray}
{\cal L}_{\tilde{q}\tilde{q}gg}&=&- \frac{g^{2}_{s}}{6} \sum_{q=u,d} \tilde{q}^{*}_{r}\tilde{q}_{r}g^{a}_{m}g^{am}-
g^{2}_{s}(d^{abc}+ \imath f^{abc})  \sum_{q=u,d} \tilde{q}^{*}_{r}T^{c}_{rt}\tilde{q}_{t}g^{a}_{m}g^{bm}.
\label{regrafeynman4}
\end{eqnarray}
$f^{abc}g^{a}_{m}g^{bm}=0$ because $f^{abc}$ is totally antisymmetric while $g^{a}_{m}g^{bm}$ 
is symmetric. The Feynman rule is represented in Fig. \ref{fig4}, in agreement with \cite{dress,tata} .

\subsubsection{Gluino-quark-squark interaction}
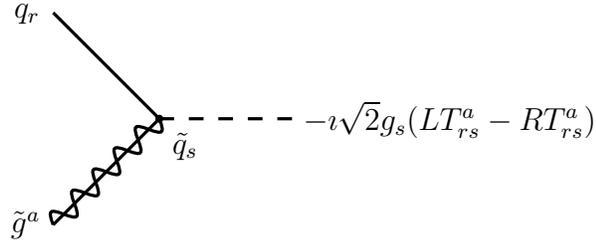
\begin{figure}[h]
\begin{center}
\begin{picture}(150,100)(0,0)
\SetWidth{1.2}
\Line(10,10)(50,50)
\Photon(10,10)(50,50){4}{5}
\Text(5,10)[r]{$\tilde{g}^{a}$}
\Line(10,90)(50,50)
\Text(5,90)[r]{$q_{r}$}
\Vertex(50,50){1.5}
\Text(60,40)[]{$\tilde{q}_{s}$}
\DashLine(100,50)(50,50){6}
\Text(160,50)[]{$- \imath \sqrt{2}g_{s}(LT^{a}_{rs}-RT^{a}_{rs})$}
\end{picture}
\end{center}
\caption{Feynman rule for the vertice gluino-quark-squark.}
\label{fig5}
\end{figure}

This interaction is described by the third line in (\ref{quarksint}), written as 
\begin{eqnarray}
{\cal L}_{\tilde{q}qg}&=&- \sqrt{2}\imath g_{s}( \bar{u}_{r}T^{a}_{rs}\tilde{u}_{s}\overline{\lambda^{a}}_{C}-
\overline{\tilde{u}}_{r}T^{a}_{rs}u_{s}\lambda^{a}_{C}+  \bar{d}_{r}T^{a}_{rs}\tilde{d}_{s}\overline{\lambda^{a}}_{C}-
\overline{\tilde{d}}_{r}T^{a}_{rs}d_{s}\lambda^{a}_{C} \nonumber \\ &+&
 \overline{u^{c}}_{r}\bar{T}^{a}_{rs}\tilde{u}^{c}_{s}\overline{\lambda^{a}}_{C}-
\overline{\tilde{u}^{c}}_{r}\bar{T}^{a}_{rs}u^{c}_{s}\lambda^{a}_{C}+
\overline{d^{c}}_{r}\bar{T}^{a}_{rs}\tilde{d}^{c}_{s}\overline{\lambda^{a}}_{C}-
\overline{\tilde{d}^{c}}_{r}\bar{T}^{a}_{rs}d^{c}_{s}\lambda^{a}_{C}).
\label{sgqsq}
\end{eqnarray}
Using the Eqs.(\ref{quarks spinors},\ref{gluino spinor}) and the usual chiral projectors
\begin{eqnarray}
L = \frac{1}{2} \left(1+ \gamma_{5} \right), \,\
R = \frac{1}{2} \left(1- \gamma_{5} \right) \,\ ,
\label{projectors}
\end{eqnarray}
we can rewrite this Lagrangian density in the following way
\begin{eqnarray}
{\cal L}_{\tilde{q}qg}&=&- \sqrt{2}g_{s} \sum_{q=u,d}( 
\bar{\Psi}( \tilde{g}^{a})L\Psi(q_{r})T^{a}_{rs}\tilde{q}^{*}_{sL}+
 \bar{\Psi}( q_{r})RT^{a}_{rs}\Psi( \tilde{g}^{a})\tilde{q}_{sL}-
 \bar{\Psi}( q_{r})LT^{a}_{rs}\Psi( \tilde{g}^{a})\tilde{q}_{sL} \nonumber \\ &-&
\bar{\Psi}( \tilde{g}^{a})R\Psi(q_{r})T^{a}_{rs}\tilde{q}^{*}_{sR}).
\label{regrafeynman5}
\end{eqnarray}
This equation gives the Feynman rule in Fig. \ref{fig5}, in agreement 
with \cite{dress,tata}.

These Feynman rules represented in Figs.\,\ref{fig1} to \ref{fig5} apply in various versions of the supersymmetric
extension of the standard model, in particular in the N/nMSSM, USSM or MSSM3RHN as well as in the MSSM.


\begin{thebibliography}{99} 

\bibitem{sg}S. L. Glashow, {\sl Nucl. Phys.}{\bf 22}, 579, (1961);
S. Weinberg, {\sl Phys. Rev. Lett.}{\bf 19}, 1264, (1967);
A. Salam in {\sl Elementary Particle Theory: Relativistic Groups
and Analyticity}, Nobel Symposium N8 (Alquivist and Wilksells, Stockolm,
1968);
S. L. Glashow, J.Iliopoulos and L.Maini,  {\sl Phys. Rev.}{\bf D 2}, 1285, (1970).

\bibitem{gl} Yu. A. Gol'fand and E.P. Likhtman, 
ZhETF Pis. Red. {\bf 13}, 452, (1971);  
[JETP Lett. {\bf 13}, 323, (1971)].


\bibitem{va} D.V. Volkov and V.P. Akulov, 
Phys. Lett. {\bf B46}, 109, (1973).


\bibitem{wz} J. Wess and B. Zumino, Nucl. Phys. {\bf B70}, 39, (1974);
Phys. Lett. {\bf B49}, 52, (1974); 
Nucl. Phys.  {\bf B78}, 1, (1974).

\bibitem{ssm} P. Fayet, Phys. Lett. {\bf B64}, 159, (1976);
{\bf B69}. 489, (1977).

\bibitem{susy}
P. Fayet and S. Ferrara, Phys. Rep. {\bf 32}, 249, (1977);\\
M. F. Sohnius, Phys. Rep. \textbf{128}, 41, (1985);\\
H. P. Nilles, Phys. Rep. \textbf{110}, 1, (1984);\\
A. B. Lahanas and D. V. Nanopoulos, Phys. Rep. \textbf{145}, 1, (1987);\\
S. J. Gates, M. Grisaru, M. Ro\v{c}ek and W. Siegel, \textit{Superspace or One Thousand
and One Lessons in Supersymmetry} (Benjamin \& Cummings, 1983);\\
P.West, \textit{Introduction to supersymmetry and supergravity}
(World Scientific, 2nd edition, 1990);\\
S.Weinberg, \textit{The quantum theory of fields. Vol. 3. Supersymmetry}
(Cambridge Univ. Press, 2000).

\bibitem{wb}J. Wess and J. Bagger, {\it Supersymmetry and Supergravity}
2nd edition, Princeton University Press, Princeton NJ, (1992).

\bibitem {dress}M. Dress, R. M. Godbole and P. Royr, \textit{Theory and
Phenomenology of Sparticles} 1st edition, World Scientific Publishing Co. Pte.
Ltd., Singapore, (2004).

\bibitem {tata}H. E. Baer and X. Tata, \textit{Weak Scale Supersymmetry} 1st
edition, Cambridge University Press, United Kindom, (2006).

\bibitem{grav} P. Fayet,  {\sl Phys. Lett.} {\bf B70}, 461, (1977).

\bibitem{sugra} P. Nath and R. Arnowitt, {\em Phys. Lett.} {\bf B56}, 177, (1975);
D. Z. Freedman, P. van Nieuwenhuizen and S. Ferrara, {\em Phys. Rev.} {\bf D13}, 3214, (1976); 
S. Deser and B. Zumino, {\em Phys. Lett.} {\bf B62}, 335, (1976); see also
"Supersymmetry", S.Ferrara, ed. (North Holland/World Scientific, Amsterdam/Singapore,
1987).

\bibitem{ABF}
U. Amaldi, W. de Boer, H. F\"{u}rstenau, Phys. Lett. \textbf{B260}, 447, (1991).


\bibitem{INO82a}K. Inoue, A. Komatsu and S. Takeshita,
{\sl Prog. Theor. Phys.} {\bf 68}, 927, (1982); {\sl Prog. Theor. Phys.} {\bf 70}, 330, (1983).

\bibitem{running}
 V. Barger, M. S. Berger and P. Ohmann, {\em Phys. Rev.} {\bf D47}, 
1093, (1993); 
 W. de Boer, R. Ehret and D. Kazakov, {\em Z. Phys.}
 {\bf C67},  647, (1995);
W. de Boer et al., {\em Z. Phys.} {\bf C71}, 415, (1996).

\bibitem{Fayet:2001xk}
  P.~Fayet, ``About the origins of the supersymmetric standard model,''
in ``Minneapolis 2000, 30 years of supersymmetry'', Nucl.\ Phys.\ Proc.\ Suppl.\  {\bf 101}, 81, (2001) [hep-ph/0107228]; \\
``About superpartners and the origins of the supersymmetric standard model,''
  in ``The Supersymmetric World - The Beginnings of the Theory'', G. Kane and M. Shifman eds. (World Sc., 2000), p.~120 [hep-ph/ 0104302];
\\
``About R-parity and the supersymmetric standard model,'' in ``The many faces of the superworld'',
Yuri Golfand Memorial Vol., M. Shifman ed. (World Sc.) p. 476 
[hep-ph/9912413].

\bibitem{Chung:2003fi} D.~J.~H.~Chung, L.~L.~Everett, G.~L.~Kane,
S.~F.~King, J.~D.~Lykken and L.~T.~Wang, 
{\sl Phys.Rept.}{\bf 407}, 1, (2005). 

\bibitem{Ibanez:wd}
  L.~E.~Iba\~nez and G.~G.~Ross,
  {\sl Phys. Lett.} {\bf B131}, 335, (1983);
B.~Pendleton and G.~G.~Ross,
  {\sl Phys. Lett.}  {\bf B98}, 291, (1981).

\bibitem{Dimopoulos:1981yj}
  S.~Dimopoulos, S.~Raby and F.~Wilczek,
  {\sl Phys. Rev.}  {\bf D24},  1681, (1981);
S.~Dimopoulos and H.~Georgi,
  {\sl Nucl. Phys.}  {\bf B193}, 150, (1981);
L.~E.~Iba\~nez and G.~G.~Ross,
  {\sl Phys. Lett.}  {\bf B105}, 439, (1981);
M.~B.~Einhorn and D.~R.~T.~Jones,
  {\sl Nucl. Phys.}  {\bf B196}, 475, (1982).

\bibitem{Kane:1992kq}
  G.~L.~Kane, C.~F.~Kolda and J.~D.~Wells,
  {\sl Phys. Rev. Lett.}  {\bf 70}, 2686, (1993);
 J.~R.~Espinosa and M.~Quiros,
  {\sl Phys. Lett.} {\bf B302}, 51, (1993).

\bibitem{lepewwg}   
LEP Electroweak Working Group, LEPEWWG/2001-01.

\bibitem{R}P. Fayet, {\sl Nucl. Phys.}  {\bf B90}, 104, (1975).

\bibitem{r1} A. Salam and J. Strathdee, {\sl Nucl. Phys.} {\bf B87}, 85, (1975).

\bibitem{barbier} R. Barbier {\it et al.}, Phys. Rept. {\bf 420}, 1, (2005)  [hep-ph/0406039].

\bibitem{moreau} G. Moreau, hep-ph/0012156.

\bibitem{Fayet} P. Fayet and J. Iliopoulos,  {\em Phys. Lett.} {\bf B51}, 461, (1974).

\bibitem{fayet79}
 P. Fayet, Phys. Lett. {\bf B84}, 416, (1979).

\bibitem{fayetF}
 P. Fayet, Phys. Lett. {\bf B58}, 67, (1975).

\bibitem{O'R} L. O'Raifeartaigh,  {\em Nucl. Phys.} {\bf B96}, 331, (1975).

\bibitem{glu} P. Fayet, {\sl Phys. Lett.} {\bf B78}, 417, (1978).

\bibitem {10}L. Girardello and M. T. Grisaru, {\sl Nucl. Phys.}{\bf B194}, 65, (1982).


\bibitem{ff} G.R. Farrar and P. Fayet, {\sl Phys. Lett.} {\bf B76}, 575, (1978).

\bibitem{ff2} G.R. Farrar and P. Fayet, {\sl Phys. Lett.} {\bf B79}, 442, (1978).

\bibitem{ff3} G.R. Farrar and P. Fayet, {\sl Phys. Lett.} {\bf B89}, 191, (1980).

\bibitem{Nilles}
 H.P. Nilles and N. Polonsky, Nucl. Phys. {\bf B499}, 33, (1997).
 T. Banks, Y. Grossman, E. Nardi and Y. Nir,  
Phys. Rev. {\bf D52}, 5319, (1995).
 F.M. Borzumati, Y. Grossman, E. Nardi and Y. Nir,  
 Phys.Lett. {\bf B384}, 123, (1996). 
 E. Nardi, Phys. Rev. {\bf D55}, 5772, (1997).

\bibitem{hall0}
L.~Hall and M.~Suzuki.
\newblock {\em Nucl. Phys.} {\bf B231},419, (1984).


\bibitem{GH}  Y. Grossman and H. Haber, {\it Phys. Rev. Lett.} 
 {\bf 78}, 3438, (1997);
{\it   Phys.Rev.} {\bf D59}, 093008, (1999);  hep-ph/9906310.

\bibitem{dreiner}H. Dreiner, hep-ph/9707435.
 G. Bhattacharyya, Nucl. Phys. Proc. Suppl. {\bf 52A}, 83, (1997) and  
 hep-ph/9709395. B. Allanach, A. Dedes and H. Dreiner, 
{\it Phys. Rev.} {\bf D60}, 075014, (1999).

\bibitem{grs79}
M.~Gell-Mann, P.~Ramond and R.~Slansky, in Supergravity, ed. by P.~van Niewenhuizen and 
D.~Freedman, North Holland, Amsterdam, 1979.

\bibitem{ms80a}
R.~N.~Mohapatra and G.~Senjanovic,
  Phys.\ Rev.\ Lett.\  {\bf 44}, 912, (1980).

\bibitem{kim} J.E. Kim and H.P. Nilles, Phys. Lett. {\bf B 138}, 150, (1984).

\bibitem{king}
S. F. King, hep-ph/9806440, CERN-TH/98-208.

\bibitem{Davidson:1998bi}
  S.~Davidson and S.~F.~King,
  Phys.\ Lett.\  B {\bf 445}, 191, (1998)
  [arXiv:hep-ph/9808296].


\bibitem{Antusch:2006bw}
  S.~Antusch, E.~Arganda, M.~J.~Herrero and A.~M.~Teixeira,
  Nucl.\ Phys.\ Proc.\ Suppl.\  {\bf 169}, 155, (2007)
  [arXiv:hep-ph/0610439].

\bibitem{Teixeira:2007gq}
  A.~M.~Teixeira, S.~Antusch, E.~Arganda and M.~J.~Herrero,
{\it Proc. 5th Flavor Physics and CP Violation Conf. (FPCP 2007), Bled, Slovenia, 12-16 May 2007, p. 029}
  [arXiv:0708.2617 [hep-ph]].
\end{thebibliography}
\end{document}